\documentclass[%
  reprint,
  showpacs,preprintnumbers,
  amsmath,amssymb,
  aps,
  prb,
]{revtex4-2}

\usepackage{graphicx}[]
\usepackage{dcolumn}
\usepackage{bm}
\usepackage{multirow}
\usepackage{braket}
\usepackage{color}
\usepackage{ulem}

\usepackage{url}

\begin{document}

\preprint{APS/123-QED}

\title{
Ferroaxial moment induced by vortex spin texture
}

\author{Satoru Hayami}
\affiliation{
Graduate School of Science, Hokkaido University, Sapporo 060-0810, Japan 
}
 
\begin{abstract}
The nature of an electric ferro-axial moment characterizing a time-reversal-even axial-vector quantity is theoretically investigated under magnetic orderings. 
We clarify that a vortex spin texture results in the emergence of the ferro-axial moment depending on the helicity. 
By introducing a multipole description, we show that the ferro-axial nature appears when two types of odd-parity magnetic multipoles become active under the vortex spin texture: One is the magnetic monopole and the other is the magnetic toroidal dipole. 
We present all the magnetic point groups to possess the ferro-axial moment in the presence of the magnetic orderings with the magnetic monopole and magnetic toroidal dipole. 
Moreover, we specifically consider a multi-orbital model in a four-sublattice tetragonal system to demonstrate the ferro-axial moment induced by the vortex spin texture. 
We show that the ferro-axial moment is microscopically characterized by both the cluster-scale and atomic-scale axial-vector quantities: The former is described by the vortex of the local electric dipole and the latter is represented by the outer product of the local orbital and spin angular momenta. 
Moreover, we find that the atomic-spin orbit coupling and the hybridization between orbitals with different parity are key ingredients to induce the ferro-axial moment.  
We also apply the result to a magnetic skyrmion with a topologically-nontrivial spin texture and discuss candidate materials. 
\end{abstract}
\maketitle

\section{Introduction}

One of the central issues in condensed matter physics is to explore intriguing functional properties in materials. 
The spontaneous symmetry breaking owing to electron correlation often gives rise to new quantum states of matter, which become the sources of various physical phenomena. 
In particular, antiferromagnetic ordering has been extensively studied in both experiments and theory over the years. 
It exhibits a variety of physical phenomena depending on its spatial distribution of spin moments. 
For example, a collinear antiferromagnetic ordering can be the origin of the anomalous Hall effect, which is termed the crystal Hall effect~\cite{Solovyev_PhysRevB.55.8060, Sivadas_PhysRevLett.117.267203, li2019quantum,vsmejkal2020crystal, feng2020observation, Shao_PhysRevApplied.15.024057, samanta2020crystal, Naka_PhysRevB.102.075112, Hayami_PhysRevB.103.L180407,lei2021large, smejkal2022anomalous, Chen_PhysRevB.106.024421}. 
Similarly, a noncoplanar antiferromagnetic ordering with a uniform distribution of the spin scalar chirality causes the anomalous Hall effect through the spin Berry phase mechanism~\cite{Nagaosa_RevModPhys.82.1539, Ohgushi_PhysRevB.62.R6065, Shindou_PhysRevLett.87.116801,taguchi2001spin,tatara2002chirality, Martin_PhysRevLett.101.156402, Neubauer_PhysRevLett.102.186602}. 
In both cases, the emergence of the anomalous Hall effect is understood from symmetry; it occurs when the magnetic point group symmetry of the antiferromagnetic ordering is the same as that of the ferromagnetic one to have the same symmetry as the axial-vector with time-reversal odd. 
In other words, antiferromagnetic orderings can acquire a ferromagnetic nature even without uniform magnetization. 

Furthermore, antiferromagnetic orderings can exhibit a polar nature when they break the spatial inversion symmetry. 
A typical example is a spontaneous electric polarization induced by a noncollinear antiferromagnetic ordering with the vector spin chirality degree of freedom~\cite{Katsura_PhysRevLett.95.057205,Mostovoy_PhysRevLett.96.067601,SergienkoPhysRevB.73.094434,Harris_PhysRevB.73.184433,Bulaevskii_PhysRevB.78.024402,ArimaJPSJ.80.052001,tokura2014multiferroics,batista2016frustration,cardias2020first,Hayami_PhysRevB.105.024413}; the ferroelectric (time-reversal-even polar) property appears below the critical temperature.
Another example is found in a noncoplanar antiferromagnetic ordering, where nonlinear nonreciprocal transport is caused by the antisymmetric band modulation under the spatial distribution of the local spin scalar chirality~\cite{Hayami_PhysRevB.101.220403, Hayami_PhysRevB.102.144441, Hayami_PhysRevResearch.3.043158, hayami2021phase, Hayami_PhysRevB.105.024413, Hayami_PhysRevB.106.014420, Hayami_doi:10.7566/JPSJ.91.094704}. 
In this case, the antiferromagnetic state accompanies the nature of ferroic time-reversal-odd polar-tensor quantities, which is referred to as ferromagnetic toroidicity. 
These examples show that antiferromagnetic spin configurations bring about various physical phenomena depending on their spin alignments, which will be utilized for an active research field in terms of antiferromagnetic spintronics~\cite{jungwirth2016antiferromagnetic, Baltz_RevModPhys.90.015005,vsmejkal2018topological,jungfleisch2018perspectives}. 

In the present study, we focus on a ferro-axial (ferro-rotational) nature in magnets. 
The ferro-axial moment is characterized by the axial vector with time-reversal even, which is qualitatively different from the ferromagnetic moment (axial vector with time-reversal odd) and ferroelectric moment (polar vector with time-reversal even)~\cite{Hlinka_PhysRevLett.116.177602,cheong2018broken}. 
Accordingly, the spatial inversion and time-reversal properties are different from each other: the former ferro-axial moment keeps both spatial inversion and time-reversal symmetries, while the latter ferromagnetic and ferroelectric moments break the time-reversal and spatial inversion symmetries, respectively. 
In this sense, the ferro-axial moment is not directly coupled to the electromagnetic fields. 
Meanwhile, such a ferro-axial property has recently attracted much attention, since it leads to unconventional off-diagonal responses, such as the spin-current generation~\cite{hayami2021electric, Roy_PhysRevMaterials.6.045004} and antisymmetric thermopolarization~\cite{Nasu_PhysRevB.105.245125}, in the context of the electric toroidal moment. 
As the ferro-axial moment can be present in the 13 crystallographic point groups without vertical mirror symmetry, $C_{\rm 6h}$, $C_{6}$, $C_{\rm 3h}$, $C_{\rm 4h}$, $C_{4}$, $S_4$, $C_{\rm 3i}$, $C_3$, $C_{\rm 2h}$, $C_2$, $C_{\rm s}$, $C_{\rm i}$, and $C_1$, a variety of materials are expected to exhibit the ferro-axial physics. 
Indeed, the ferro-axial nature has been experimentally observed in materials, such as Co$_3$Nb$_2$O$_8$~\cite{Johnson_PhysRevLett.107.137205}, CaMn$_7$O$_{12}$~\cite{Johnson_PhysRevLett.108.067201}, RbFe(MoO$_4$)$_2$~\cite{jin2020observation,Hayashida_PhysRevMaterials.5.124409}, NiTiO$_3$~\cite{hayashida2020visualization, Hayashida_PhysRevMaterials.5.124409}, Ca$_5$Ir$_3$O$_{12}$~\cite{hanate2021first}, and BaCoSiO$_4$~\cite{Xu_PhysRevB.105.184407}. 

Motivated by these studies, we study the possibility of inducing and controlling such a ferro-axial nature through magnetic phase transitions. 
By introducing an odd-parity multipole notation~\cite{hayami2018microscopic, Hayami_PhysRevB.98.165110,kusunose2020complete, Yatsushiro_PhysRevB.104.054412}, we reveal that the coexistence of the magnetic monopole and the magnetic toroidal dipole under the vortex spin texture is the key essence to inducing the ferro-axial nature in antiferromagnets. 
We summarize possible magnetic point groups to possess the ferro-axial moment under antiferromagnetic orderings with the above two multipoles. 
Furthermore, we identify the electronic degrees of freedom corresponding to the ferro-axial moment, which is defined in a cluster and atomic site, and show when a nonzero expectation value is obtained based on the microscopic model. 
We show that the atomic spin-orbit coupling and the hybridization between orbitals with different parity are key ingredients in addition to the vortex spin configuration. 
We also discuss a magnetic skyrmion as a canonical magnetic state to have the vortex spin configuration, which becomes the candidate to exhibit the ferro-axial physics. 

The rest of this paper is organized as follows. 
In Sec.~\ref{sec: Ferroaxial moment under vortex spin textures}, we introduce the ferro-axial moment based on the multipole description.  
We show that the ferro-axial moment corresponds to the electric toroidal dipole. 
In addition, we discuss two key multipoles, i.e., magnetic monopole and magnetic toroidal dipole, to induce the ferro-axial moment under antiferromagnetic orderings. 
We also present the reduction table from the gray point group to the subgroups once the ferro-axial moment becomes active in the presence of the magnetic monopole and the magnetic toroidal dipole. 
Then, after introducing a minimum multi-orbital model in Sec.~\ref{sec: Model}, we demonstrate that the ferro-axial moment is induced by the vortex spin textures in Sec.~\ref{sec: Result}. 
We show that the interplay between the atomic spin-orbit coupling and the hybridization between orbitals with different parity plays an important role in inducing the ferro-axial nature under the vortex spin configuration. 
In Sec.~\ref{sec: Relevance with skyrmion}, we show that the magnetic skyrmion is one of the candidates to exhibit the ferro-axial nature. 
Section~\ref{sec: Summary} concludes this paper. We list the candidate materials. 
In Appendix~\ref{sec: Ferro-axial moment under anti-vortex spin textures}, we show that a superposition of the anti-vortices with the magnetic quadrupole degrees of freedom also leads to the emergence of the ferro-axial moment. 
In Appendix~\ref{sec: Filling dependence of ferro-axial moment}, we discuss the filling dependences of the ferro-axial moment in the model in Sec.~\ref{sec: Model}.

\section{Ferro-axial moment under vortex spin textures}
\label{sec: Ferroaxial moment under vortex spin textures}

We discuss an essence to induce the ferro-axial moment under antiferromagnetic orderings based on the microscopic multipole description. 
First, we introduce four types of multipoles with different spatial inversion and time-reversal properties in Sec.~\ref{sec: Ferroaxial moment based on multipole description}. 
We show that the ferro-axial moment corresponds to the electric toroidal dipole degree of freedom in electrons. 
Then, we show that the ferro-axial nature appears under the vortex spin texture accompanying both the magnetic monopole and magnetic toroidal dipole degrees of freedom in Sec.~\ref{sec: Vortex spin configuration}. 
Lastly, we list the comprehensive table to represent the reduction from the gray point group to the subgroups with a nonzero ferro-axial moment in Sec.~\ref{sec: Magnetic point group}.

\subsection{Ferro-axial moment based on multipole description}
\label{sec: Ferroaxial moment based on multipole description}

\begin{table}[htb!]
\centering
\caption{
Parities for eight multipoles $(Q_0, G_z, M_0, T_z, G_0, Q_z, T_0, M_z)$ in terms of the symmetry operations under the $\infty/mm 1'$ ($D'_{\infty h}$) group with the irreducible representation (Irrep.); the upper subscript corresponds to the time-reversal parity.  
$E$, $\bar{1}$, $m_{\parallel}$, and $2_{\perp}$ represent the symmetries in terms of identity, spatial inversion, mirror parallel to the principal axis, and twofold rotation around the principal axis, respectively. 
The prime symbol for $E$, $\bar{1}$, $m_{\parallel}$, and $2_{\perp}$ stands for the operations combined with time reversal. 
In the rightmost column, symmetry equivalent expressions by using the position vector $\bm{r}$ and the dipoles for $\bm{X}=\bm{Q}$, $\bm{M}$, $\bm{G}$, and $\bm{T}$ are shown. 
\label{table: symmetry}}
\begin{tabular}{ccccccccccccc}\hline \hline
Multipole & $E$  & $\bar{1}$ & $m_{\parallel}$  & $2_{\perp}$ & $1'$  & $\bar{1}'$ & $m'_{\parallel}$  & $2'_{\perp}$ & Irrep. & Remark \\
\hline
$Q_0$ & $+1$ & $+1$ & $+1$ & $+1$ & $+1$ & $+1$ & $+1$ & $+1$ & $A_{1g}^+$ & $\bm{r} \cdot \bm{Q}$ \\ 
$G_z$ & $+1$ & $+1$ & $-1$ & $-1$ & $+1$ & $+1$ & $-1$ & $-1$ & $A_{2g}^+$ & $(\bm{r} \times \bm{Q})_z$ \\ 
\hline
$M_0$ & $+1$ & $-1$ & $-1$ & $+1$ & $-1$ & $+1$ & $+1$ & $-1$ & $A^-_{2u}$ & $\bm{r} \cdot \bm{M}$ \\
$T_z$  & $+1$ & $-1$ & $+1$ & $-1$ & $-1$ & $+1$ & $-1$  & $+1$ & $A_{1u}^-$ & $(\bm{r} \times \bm{M})_z$\\
\hline
$G_0$ & $+1$ & $-1$ & $-1$ & $+1$ & $+1$ & $-1$ & $-1$ & $+1$ & $A_{2u}^+$ & $\bm{r} \cdot \bm{G}$\\ 
$Q_z$ & $+1$ & $-1$ & $+1$ & $-1$ & $+1$ & $-1$ & $+1$ & $-1$ & $A_{1u}^+$ & $(\bm{r} \times \bm{G})_z$\\ 
\hline
$T_0$ & $+1$ & $+1$ & $+1$ & $+1$ & $-1$ & $-1$ & $-1$ & $-1$ & $A^-_{1g}$ & $\bm{r} \cdot \bm{T}$\\
$M_z$  & $+1$ & $+1$ & $-1$ & $-1$ & $-1$ & $-1$ & $+1$  & $+1$ & $A_{2g}^-$ & $(\bm{r} \times \bm{T})_z$
 \\
\hline\hline
\end{tabular}
\end{table}

According to the spatial inversion ($\mathcal{P}$ or $\bar{1}$) and time-reversal ($\mathcal{T}$ or $1'$) properties, there are four fundamental multipole degrees of freedom: electric, magnetic, electric toroidal, and magnetic toroidal multipoles~\cite{zel1958relation,dubovik1975multipole,dubovik1990toroid,hayami2018microscopic,Hayami_PhysRevB.98.165110,kusunose2020complete,Yatsushiro_PhysRevB.104.054412}. 
For example, the monopole component of electric ($Q_0$), magnetic ($M_0$), electric toroidal ($G_0$), and magnetic toroidal ($T_0$) is characterized by $(\mathcal{P}, \mathcal{T})=(+1,+1)$, $(-1, -1)$, $(-1,+1)$, and $(+1,-1)$, respectively, while the dipole component of electric [$\bm{Q}=(Q_x,Q_y,Q_z)$], magnetic [$\bm{M}=(M_x,M_y,M_z)$], electric toroidal [$\bm{G}=(G_x,G_y,G_z)$], and magnetic toroidal [$\bm{T}=(T_x,T_y,T_z)$] is characterized by $(\mathcal{P}, \mathcal{T})=(-1,+1)$, $(+1, -1)$, $(+1,+1)$, and $(-1,-1)$, respectively. 
Although the parities of $Q_0$, $M_0$, $G_0$, and $T_0$ in terms of $\mathcal{P}$ and $\mathcal{T}$ are the same as $\bm{G}$, $\bm{T}$, $\bm{Q}$, and $\bm{M}$, respectively, their transformation in terms of the point-group symmetry operations are different from each other. 
To demonstrate that, we show parities for these eight multipoles in terms of the symmetry operations under the $\infty/mm 1'$ ($D'_{\infty h}$) group in Table~\ref{table: symmetry}~\cite{Hlinka_PhysRevLett.113.165502}. 
We also denote the irreducible representation under $\infty/mm 1'$ where the superscript stands for the time-reversal parity. 
The eight multipoles belong to the different irreducible representations, which means that they are independent under the $\infty/mm 1'$ group. 
Such a classification of multipoles can be performed for an arbitrary point group~\cite{Hayami_PhysRevB.98.165110, Yatsushiro_PhysRevB.104.054412}. 

Among the eight multipoles, the electric ($Q_0$ and $\bm{Q}$) and magnetic toroidal ($T_0$ and $\bm{T}$) multipoles are characterized by the polar-tensor quantity, while the magnetic ($M_0$ and $\bm{M}$) and electric toroidal ($G_0$ and $\bm{G}$) multipoles are characterized by the axial-tensor quantity. 
These multipoles are transformed from each other by taking the inner (outer) product of any dipoles $\bm{X}=(\bm{Q}, \bm{M}, \bm{G}, \bm{T})$ and the position vector (time-reversal-even polar vector) $\bm{r}$; $\bm{r} \cdot \bm{X}$ is symmetrically equivalent to $X_0$ and $\bm{r} \times \bm{X}$ results in the change of the spatial inversion parity while keeping its rank. 
For example, $\bm{r} \cdot \bm{Q}$ corresponds to $Q_0$ and $\bm{r} \times \bm{Q}$ corresponds to $\bm{G}$~\cite{dubovik1990toroid,naumov2004unusual,Prosandeev_PhysRevB.75.094102}. 
In other words, nonzero $\bm{\nabla}\cdot \bm{Q}_{\parallel}$ but $\bm{\nabla}\cdot \bm{Q}_{\perp}=0$ corresponds to $Q_0$, while nonzero $\bm{\nabla}\times \bm{Q}_{\perp}$ but $\bm{\nabla}\times \bm{Q}_{\parallel}=0$ correspond to $\bm{G}$ when we decompose $\bm{Q}$ into the perpendicular component $\bm{Q}_{\perp}$ and parallel one $\bm{Q}_{\parallel}$~\cite{dubovik1990toroid}.
We present such correspondence in Table~\ref{table: symmetry}.

Since only four types of dipole degrees of freedom are present according to the spatial inversion and time-reversal properties, any vector physical quantities are represented by ($\bm{Q}, \bm{M}, \bm{G}, \bm{T}$). 
For example, the electric polarization (ferroelectric moment), the magnetization (ferromagnetic moment), and the electric current (ferro-magnetic-toroidal moment) have a correspondence to $\bm{Q}$, $\bm{M}$, and $\bm{T}$, respectively, from the symmetry viewpoint. 
In addition, the ferro-axial moment, which is a time-reversal-even axial-vector quantity, corresponds to the electric toroidal dipole ($\bm{G}$), where the symmetries in terms of mirror parallel to the principal axis and twofold rotation around the principal axis are broken while keeping both the spatial inversion and time-reversal symmetries. 
Thus, in order to induce the ferro-axial nature in the electronic systems, it is important to activate the electric toroidal dipole $\bm{G}$. 
It is noted that $\bm{G}$ is distinguished from $\bm{M}$ with respect to the time-reversal parity, although both dipoles are characterized by the axial vector. 

\subsection{Vortex spin configuration}
\label{sec: Vortex spin configuration}

\begin{figure}[t!]
\begin{center}
\includegraphics[width=1.0 \hsize ]{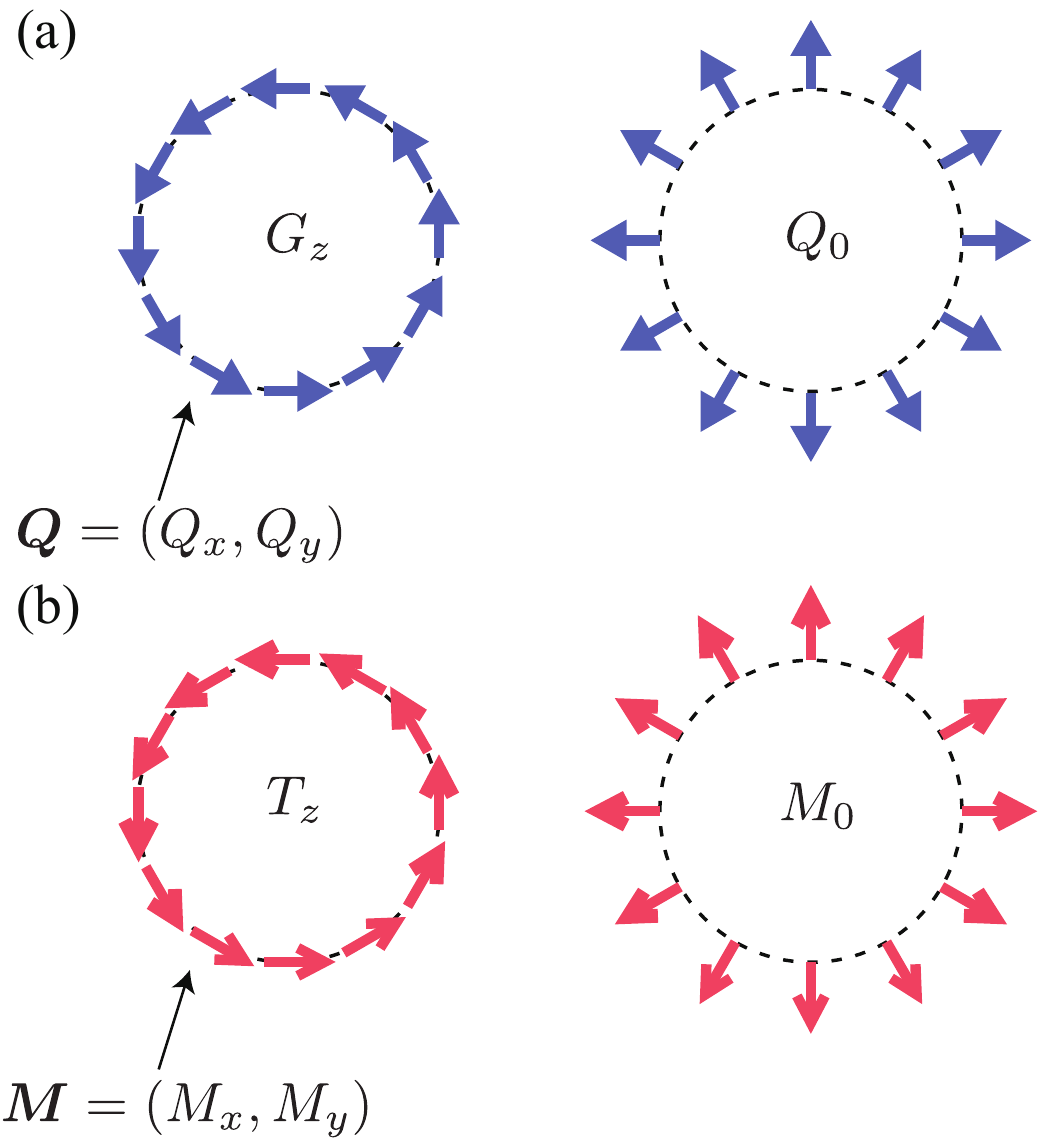} 
\caption{
\label{fig: ponti}
Vortex configurations of (a) the electric dipole $\bm{Q}=(Q_x, Q_y,0)$ and (b) the magnetic dipole $\bm{M}=(M_x, M_y,0)$. 
In (a) [(b)], the left panel shows the electric (magnetic) toroidal dipole $G_z$ ($T_z$) and the right panel shows the electric (magnetic) monopole $Q_0$ ($M_0$); $G_z$ corresponds to the ferro-axial moment. 
}
\end{center}
\end{figure}

As the electric toroidal dipole $\bm{G}$ is related to the electric dipole $\bm{Q}$ like $\bm{G} \leftrightarrow \bm{r} \times \bm{Q}$ in Table~\ref{table: symmetry}, one finds that the vortex-type alignment of the electric dipole $\bm{Q}$ can induce the ferro-axial moment. 
We show the example of the vortex to have nonzero $(\bm{r} \times \bm{Q})_z $ in the left panel of Fig.~\ref{fig: ponti}(a), where the center of the vortex is taken at the origin of the position vector, i.e., $\bm{r}=\bm{0}$. 
In such a situation, a nonzero expectation value of $G_z$ is expected in the electron system. 
Meanwhile, when the direction of $\bm{Q}$ is locally rotated so as to have the component of $\bm{r} \cdot \bm{Q}$ shown in the right panel of Fig.~\ref{fig: ponti}(a), the electric monopole $Q_0$ is activated instead of $G_z$.  
These vortices with $G_z$ and $Q_0$ in Fig.~\ref{fig: ponti}(a) show the same symmetry property shown in Table~\ref{table: symmetry}; $G_z$ belongs to $A_{2g}^+$ and $Q_0$ belongs to $A_{1g}^+$ under the $\infty /mm 1'$ group. 

In a similar manner, the vortex-type alignment of the magnetic dipole (spin) $\bm{M}$ leads to different multipoles. 
The magnetic toroidal dipole $T_z$ is activated for the vortex spin configuration with the component of $\bm{r} \times \bm{M}$, while the magnetic monopole $M_0$ is activated for that with the component of $\bm{r} \cdot \bm{M}$, as shown in Fig.~\ref{fig: ponti}(b). 
The irreducible representations of $T_z$ and $M_0$ are represented by $A^-_{1u}$ and $A^-_{2u}$ under the $\infty /mm 1'$ group, respectively. 
$T_z$ and $M_0$ correspond to the odd-parity magnetic multipoles~\cite{Spaldin_0953-8984-20-43-434203,kopaev2009toroidal,cheong2018broken,Gao_PhysRevB.97.134423}, which are the sources of the linear magnetoelectric effect~\cite{popov1999magnetic,arima2005resonant,van2007observation,Yanase_JPSJ.83.014703,zimmermann2014ferroic,Hayami_PhysRevB.90.024432,Hayami_doi:10.7566/JPSJ.84.064717,Toledano_PhysRevB.92.094431,yatsushiro2019atomic,Hayami_PhysRevB.104.045117}, nonlinear (spin) current generation~\cite{Sawada_PhysRevLett.95.237402,Kezsmarki_PhysRevLett.106.057403,Miyahara_JPSJ.81.023712,Miyahara_PhysRevB.89.195145,Bordacs_PhysRevB.92.214441,Sato_PhysRevLett.124.217402,Yatsushiro_PhysRevB.105.155157,Kondo_PhysRevResearch.4.013186,Hayami_PhysRevB.106.024405}, and nonreciprocal magnon excitations~\cite{Iguchi_PhysRevB.92.184419,Hayami_doi:10.7566/JPSJ.85.053705,Gitgeatpong_PhysRevLett.119.047201,sato2019nonreciprocal,Matsumoto_PhysRevB.101.224419,Matsumoto_PhysRevB.104.134420,Hayami_PhysRevB.105.014404}. 
Similarly, one expects that $G_0$ and $Q_z$ ($T_0$ and $M_z$) are related to the vortices of $\bm{G}$ ($\bm{T}$), as shown in Table~\ref{table: symmetry}, although we here do not consider them, since we focus on the ferro-axial nature induced by the vortex spin textures in Fig.~\ref{fig: ponti}(b).

\begin{figure}[htb!]
\begin{center}
\includegraphics[width=1.0 \hsize ]{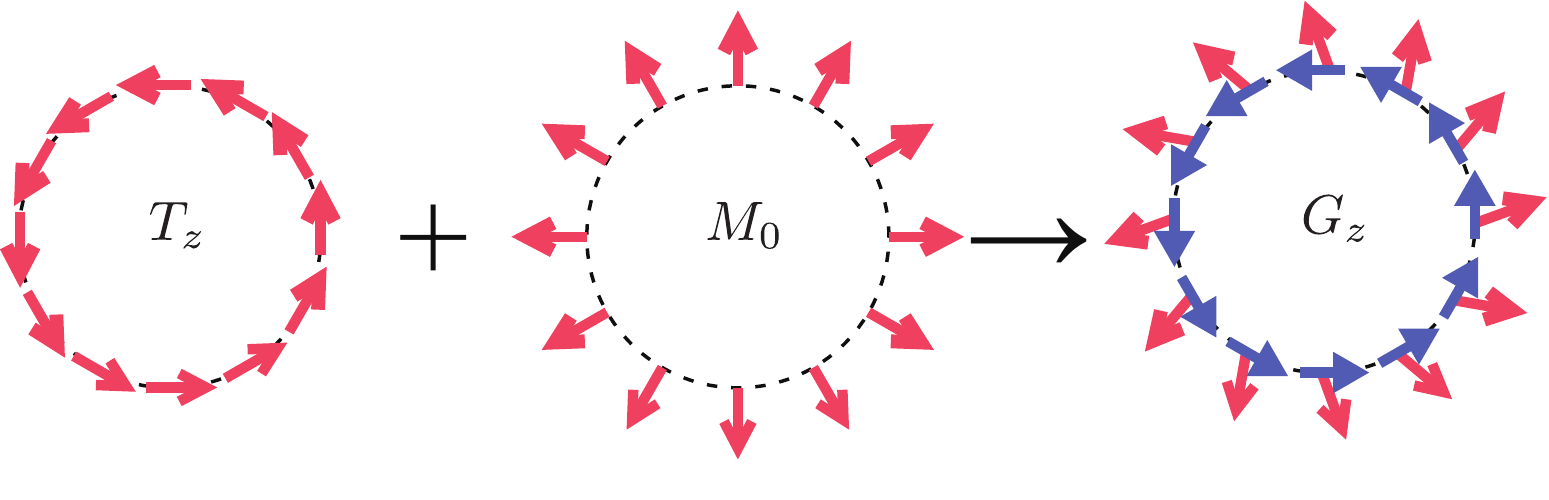} 
\caption{
\label{fig: ponti2}
The ferro-axial moment ($G_z$) induced by the superposition of the spin vortices with $T_z$ and $M_0$. 
}
\end{center}
\end{figure}

The individual vortex spin configuration in Fig.~\ref{fig: ponti}(b) does not induce the ferro-axial nature because $G_z$ belongs to the different irreducible representation of $T_{z}$ and $M_0$. 
Meanwhile, when considering a superposition of the vortices with $T_z$ and $M_0$, one can realize the activation of $G_z$ owing to a further symmetry lowering, as shown in Fig.~\ref{fig: ponti2}. 
In this situation, an effective coupling of $M_0 T_z$ occurs by the secondary effect of two order-parameter components $M_0$ and $T_z$.
Since the transformation property of $M_0 T_z$ with respect to the symmetry operations is the same as that of $G_z$ as presented in Table~\ref{table: symmetry}, the vortex spin configurations with both $T_z$ and $M_0$ exhibit a ferro-axial nature. 
In this way, the ferro-axial moment can be driven by magnetic orderings with $T_z$ and $M_0$; the onset of $G_z$ is caused by the magnetic phase transitions.

\subsection{Magnetic point group}
\label{sec: Magnetic point group}

\begin{table}[htb!]
\centering
\caption{
Reduction from the gray point group (GPG) to the subgroups when nonzero $M_0$, $T_z$, and $G_z$ appear. 
The other active multipoles, $M_z$, $G_0$, $Q_z$, and $T_0$, are also presented by $\checkmark$. 
It is noted that $Q_0$ is always active, since it belongs to the totally symmetry representation. 
\label{table: MPG}}
\begin{tabular}{ccccccccccccc}\hline \hline
GPG & $M_0, T_z, G_z$  & $M_z$ & $G_0$ & $Q_z$ & $T_0$\\
\hline 
$6/mmm 1'$, $6/m 1'$ & $6/m'$ & -- & -- & --  & -- \\
$622 1'$, $6mm 1'$, $6 1'$ & $6$ & $\checkmark$ & $\checkmark$ & $\checkmark$ & $\checkmark$ \\
$\bar{6}m2 1'$, $\bar{6} 1'$ & $\bar{6}'$ & -- & -- & -- & -- \\ \hline
$\bar{3}m 1'$, $\bar{3} 1'$ & $\bar{3}'$ & -- & -- & -- & --  \\
$32 1'$, $3m 1'$, $3 1'$ & $3$ & $\checkmark$ & $\checkmark$ & $\checkmark$ & $\checkmark$\\ \hline
$4/mmm 1'$, $4/m 1'$  & $4/m'$ & -- & -- & -- & -- \\
$422 1'$, $4mm 1'$, $4 1'$ & $4$ & $\checkmark$ & $\checkmark$ & $\checkmark$ & $\checkmark$ \\
$\bar{4}2m 1'$, $\bar{4}1'$ & $\bar{4}'$ & -- & -- & -- & --  \\ \hline
$mmm 1'$, $2/m 1'$ & $2/m'$ & -- & -- & -- & --  \\
$222 1'$, $mm2 1'$, $2 1'$ & $2$ & $\checkmark$ & $\checkmark$ & $\checkmark$ & $\checkmark$\\ 
$m 1'$ & $m'$ & --\footnote{Nonzero $(M_x, M_y)$.} & -- & --\footnote{Nonzero $(Q_x, Q_y)$.} & --  \\ \hline
$\bar{1} 1'$ & $\bar{1}'$ & -- & -- & -- & --  \\
$1 1'$ & $1$ & $\checkmark^{\rm a}$ & $\checkmark$ & $\checkmark^{\rm b}$ & $\checkmark$
 \\
\hline\hline
\end{tabular}
\end{table}

The above argument of the symmetry correspondence between $G_z$ and $M_0 T_z$ holds for any point group~\cite{Yatsushiro_PhysRevB.104.054412}. 
We show the symmetry reduction from the gray point group except for the cubic point group to the subgroups with nonzero $M_0$, $T_z$, and $G_z$ in Table~\ref{table: MPG}. 
As shown in Table~\ref{table: MPG}, 13 out of 122 magnetic point groups satisfy the condition to possess $M_0$, $T_z$, and $G_z$, which are reduced from 32 gray point groups with the time-reversal symmetry~\cite{comment_ET}. 
It is noted that there are several magnetic point groups to activate $G_z$ under the anti-vortex spin configurations with the magnetic quadrupole degrees of freedom instead of $M_0$ and $G_z$, as discussed in Appendix~\ref{sec: Ferro-axial moment under anti-vortex spin textures}. 
In the following section, we demonstrate that $G_z$ is induced in the vortex spin configuration with $M_0$ and $T_z$ by considering a specific lattice system so that the symmetry reduction occurs from $4/mmm 1'$ to $4/m'$.

\section{Model}
\label{sec: Model}

\begin{figure}[t!]
\begin{center}
\includegraphics[width=1.0 \hsize ]{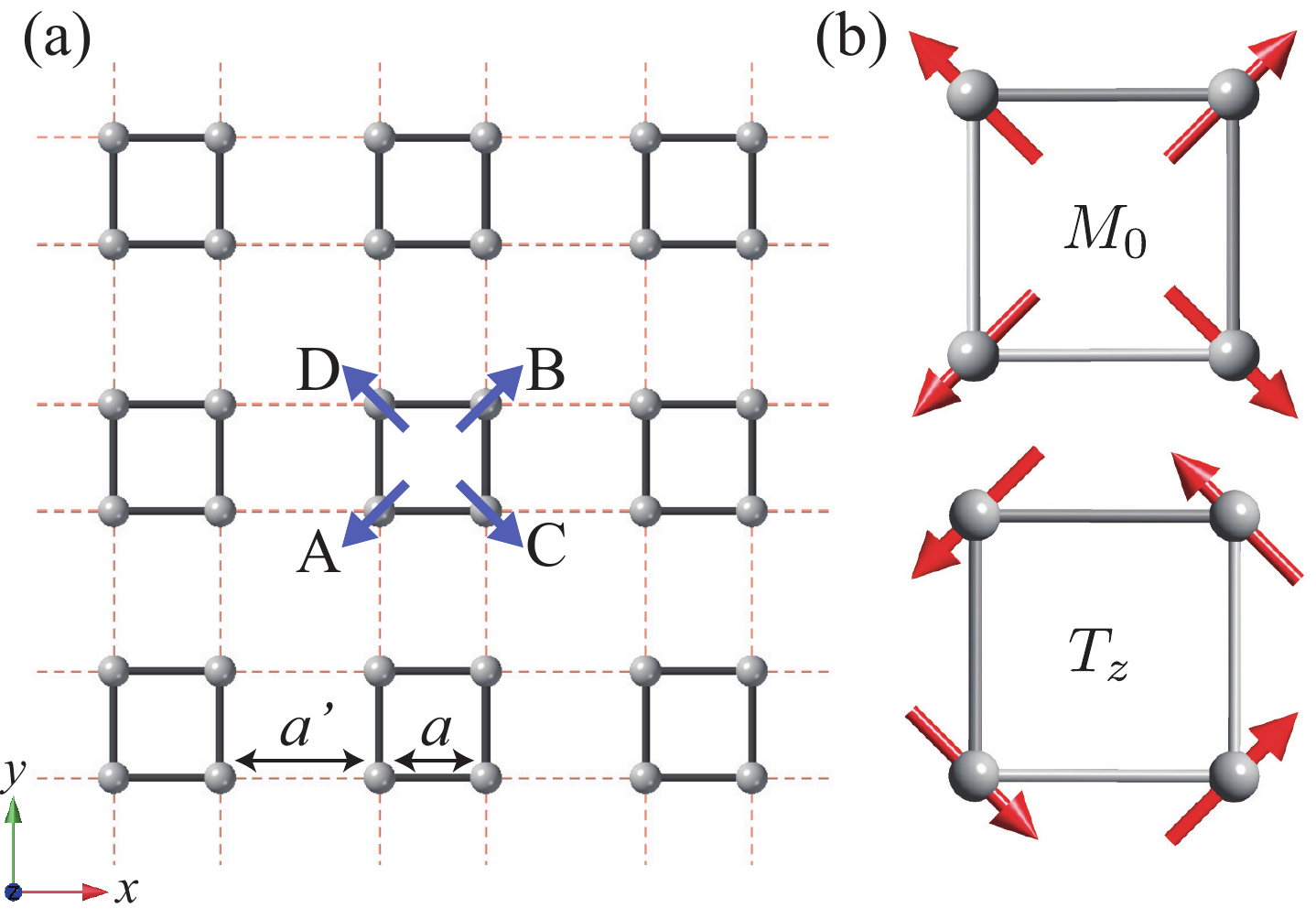} 
\caption{
\label{fig: Lattice}
(a) Two-dimensional tetragonal lattice structure consisting of four sublattices A--D. 
The blue arrows represent the local electric field. 
(b) The four-sublattice spin configurations of the magnetic monopole $M_0$ (upper panel) and the magnetic toroidal dipole $T_z$ (lower panel).
}
\end{center}
\end{figure}

Let us consider a two-dimensional tetragonal system consisting of four sublattices A-D under the magnetic point group $4/mmm 1'$ in Fig.~\ref{fig: Lattice}(a). 
We also take into account the four orbital degrees of freedom, $s$, $p_x$, $p_y$, and $p_z$, at each site in order to activate the local electric dipole degree of freedom, as discussed below. 
Then, the model Hamiltonian is given by 
\begin{align}
\label{eq: Ham_multisite}
\mathcal{H}=\sum_{\bm{k}} \sum_{\gamma,\gamma'}\sum_{\alpha,\alpha'} \sum_{\sigma,\sigma'}c^{\dagger}_{\bm{k}\gamma \alpha \sigma} H^{\gamma\gamma'}_{\alpha\alpha'\sigma\sigma'} c_{\bm{k}\gamma' \alpha'\sigma'}, 
\end{align}
where $c^{\dagger}_{\bm{k}\gamma \alpha \sigma}$ ($c_{\bm{k}\gamma \alpha \sigma}$) is the creation (annihilation) operator of electrons at wave vector $\bm{k}$, sublattice $\gamma=$ A-D, orbital $\alpha=s$, $p_x$, $p_y$, and $p_z$, and spin $\sigma$. 
The Hamiltonian matrix is divided into three parts as follows: 
\begin{align}
\label{eq: Ham_mat}
H^{\gamma\gamma'}_{\alpha\alpha'\sigma\sigma'}= \delta_{\sigma\sigma'}H^{{\rm hop}}
+
\delta_{\sigma\sigma'} \delta_{\gamma\gamma'} 
H^{{\rm SOC}}
+
\delta_{\alpha\alpha'}H^{{\rm MF}}. 
\end{align}
The first term $H^{\rm hop}$ in Eq.~(\ref{eq: Ham_mat}) represents the nearest-neighbor hopping term for the intra- and inter-plaquette. 
By using the Slater-Koster parameter, we consider the four intra-plaquette hopping parameters to satisfy the tetragonal symmetry: $t$ for the amplitude between $s$ orbitals ($\alpha,\alpha'=s$), $t_p$ for that between $(p_x, p_y)$ orbitals ($\alpha,\alpha'=p_x, p_y$), $t_z$ for that between $p_z$ orbitals ($\alpha,\alpha'=p_z$), and $t_{sp}$ for that between different $s$-$(p_x, p_y)$ orbitals ($\alpha,\alpha'=s, p_x, p_y$ and $\alpha \neq \alpha'$). 
Similarly, we set the inter-plaquette hoppings as $t'$, $t'_p$, $t'_z$, and $t'_{sp}$. 
For the lattice constant, we take $a=a'=1/2$ in Fig.~\ref{fig: Lattice}(a); the difference between $a$ and $a'$ is expressed as the different hopping amplitudes. 
We also set $t=-1$ as the energy unit of the model in Eq.~(\ref{eq: Ham_multisite}) and $(t', t'_p, t'_z, t'_{sp})=\Gamma (t, t_p, t_z, t_{sp})$ for simplicity. 
The second term $H^{\rm SOC}$ in Eq.~(\ref{eq: Ham_mat}) represents the atomic spin-orbit coupling for the $p$ orbital with the amplitude $\lambda$, which divides the six $p$-orbital levels into a doublet and a quartet. 
The third term $H^{\rm MF}$ in Eq.~(\ref{eq: Ham_mat}) represents the mean-field term corresponding to the magnetic order. 
We here suppose the same coupling constant for all the orbitals for simplicity. 
In the following, we only consider two types of magnetic textures with $M_0$ and $T_z$, as shown in Fig.~\ref{fig: Lattice}(b), which correspond to the vortex spin textures in the right and left panels of Fig.~\ref{fig: ponti}(b), respectively. 
The expression of the mean-field Hamiltonian matrix is represented by 
\begin{align}
H^{\rm MF}=&
-h_{\rm M} \sum_{\gamma''} \delta_{\gamma \gamma''}\delta_{\gamma' \gamma''}(\bm{e}_{\gamma''}\cdot \bm{\sigma}_{\sigma\sigma'}) \nonumber \\
&-
h_{\rm MT} \sum_{\gamma''} \delta_{\gamma \gamma''}\delta_{\gamma' \gamma''}(\bm{e}_{\gamma''}\times \bm{\sigma}_{\sigma\sigma'}), 
\end{align}
where $\bm{\sigma}$ is the vector of the Pauli matrices and $\bm{e}_{\rm A}=(-1,-1)$, $\bm{e}_{\rm B}=(1,1)$, $\bm{e}_{\rm C}=(1,-1)$, and $\bm{e}_{\rm D}=(-1,1)$.
The first term represents the mean field for the magnetic-monopole moment and the second term represents that for the magnetic-toroidal-dipole moment in Fig.~\ref{fig: Lattice}(b). 

The model in Eq.~(\ref{eq: Ham_multisite}) corresponds to a minimum lattice model to induce the ferro-axial moment under the vortex spin texture. 
In other words, the model has a degree of freedom to describe the vortex configurations of both the electric dipole $\bm{Q}$ and the magnetic dipole $\bm{M}$ in Figs.~\ref{fig: ponti}(a) and \ref{fig: ponti}(b). 
The local electric dipole degree of freedom is expressed as the local $s$-$p$ hybridization; the local electric polarization $\bm{Q}_\gamma$ is defined by
\begin{align}
\label{eq: localQ}
\bm{Q}_\gamma= \frac{1}{N}\sum_{\bm{k}\sigma}c^{\dagger}_{\bm{k}\gamma s \sigma}c^{}_{\bm{k}\gamma \bm{p} \sigma}+{\rm h.c.}, 
\end{align}
where $N$ is the number of supercells in the system. 
The local spin moment is given as 
\begin{align}
\label{eq: localM}
\bm{M}_\gamma= \frac{1}{N}\sum_{\bm{k}\alpha\sigma\sigma'}c^{\dagger}_{\bm{k}\gamma \alpha \sigma} \bm{\sigma}_{\sigma\sigma'} c^{}_{\bm{k}\gamma \alpha \sigma'}.  
\end{align}

By using $\bm{Q}_\gamma$ and $\bm{M}_\gamma$, one evaluates the expectation values of $Q_0$, $G_z$, $M_0$, and $T_z$ induced by $\bm{Q}_\gamma$ and $\bm{M}_\gamma$, which are given in the cluster form as 
\begin{align}
\label{eq:Q0c}
\langle Q^{\rm (c)}_0 \rangle &= \frac{1}{4}\sum_{\gamma} \bm{e}_\gamma \cdot \langle\bm{Q}_\gamma\rangle, \\
\label{eq:Gzc}
\langle G^{\rm (c)}_z \rangle &= \frac{1}{4}\sum_{\gamma} \bm{e}_\gamma \times \langle\bm{Q}_\gamma\rangle, \\
\label{eq:M0c}
\langle M^{\rm (c)}_0 \rangle &= \frac{1}{4}\sum_{\gamma} \bm{e}_\gamma \cdot \langle\bm{M}_\gamma\rangle, \\
\label{eq:Tzc}
\langle T^{\rm (c)}_z \rangle &= \frac{1}{4}\sum_{\gamma} \bm{e}_\gamma \times \langle\bm{M}_\gamma\rangle, 
\end{align}
where $\langle \cdots \rangle$ represents the statistical average and we denote the superscript (c) to explicitly represent the quantity defined by the four-sublattice cluster, which is referred to as the cluster multipole~\cite{hayami2016emergent, Suzuki_PhysRevB.95.094406,suzuki2018first, Suzuki_PhysRevB.99.174407}. 
Although there is ambiguity in terms of the choice of the origin of the position vectors in Eqs.~(\ref{eq:Q0c})-(\ref{eq:Tzc}), the following qualitative feature holds for the different choice of the origin~\cite{EdererPhysRevB.76.214404, prosandeev2009hypertoroidal}. 
The definition of $G^{\rm (c)}_z$ consisting of the vortex in terms of $\langle\bm{Q}_\gamma\rangle$ corresponds to that discussed in ferroelectric systems~\cite{naumov2004unusual,Prosandeev_PhysRevB.75.094102}

From the symmetry viewpoint, $Q^{\rm (c)}_0$, $G^{\rm (c)}_z$, $M^{\rm (c)}_0$, and $T^{\rm (c)}_z$ belong to the irreducible representation of $A^+_{1g}$, $A^+_{2g}$, $A^-_{1u}$, and $A^-_{2u}$ under the $4/mmm 1'$ group, respectively. 
When the vortex spin configuration with $\langle M^{\rm (c)}_0 \rangle$ ($\langle T^{\rm (c)}_z \rangle$) occurs, the $4/mmm 1'$ group reduces to the $4/m'm'm'$ ($4/m'mm$) group, where $\langle G^{\rm (c)}_z \rangle$ does not belong to the totally symmetric representation. 
By considering a further symmetry reduction to have both $\langle M^{\rm (c)}_0 \rangle$ and $\langle T^{\rm (c)}_z \rangle$, the system reduces to the $4/m'$ group, which results in the activation of $\langle G^{\rm (c)}_z \rangle$ as shown in Table~\ref{table: MPG}. 
In the following section, we examine the model in Eq.~(\ref{eq: Ham_multisite}) to understand the microscopic origin of the ferro-axial moment beyond the symmetry argument.

\begin{figure*}[htb!]
\begin{center}
\includegraphics[width=1.0 \hsize ]{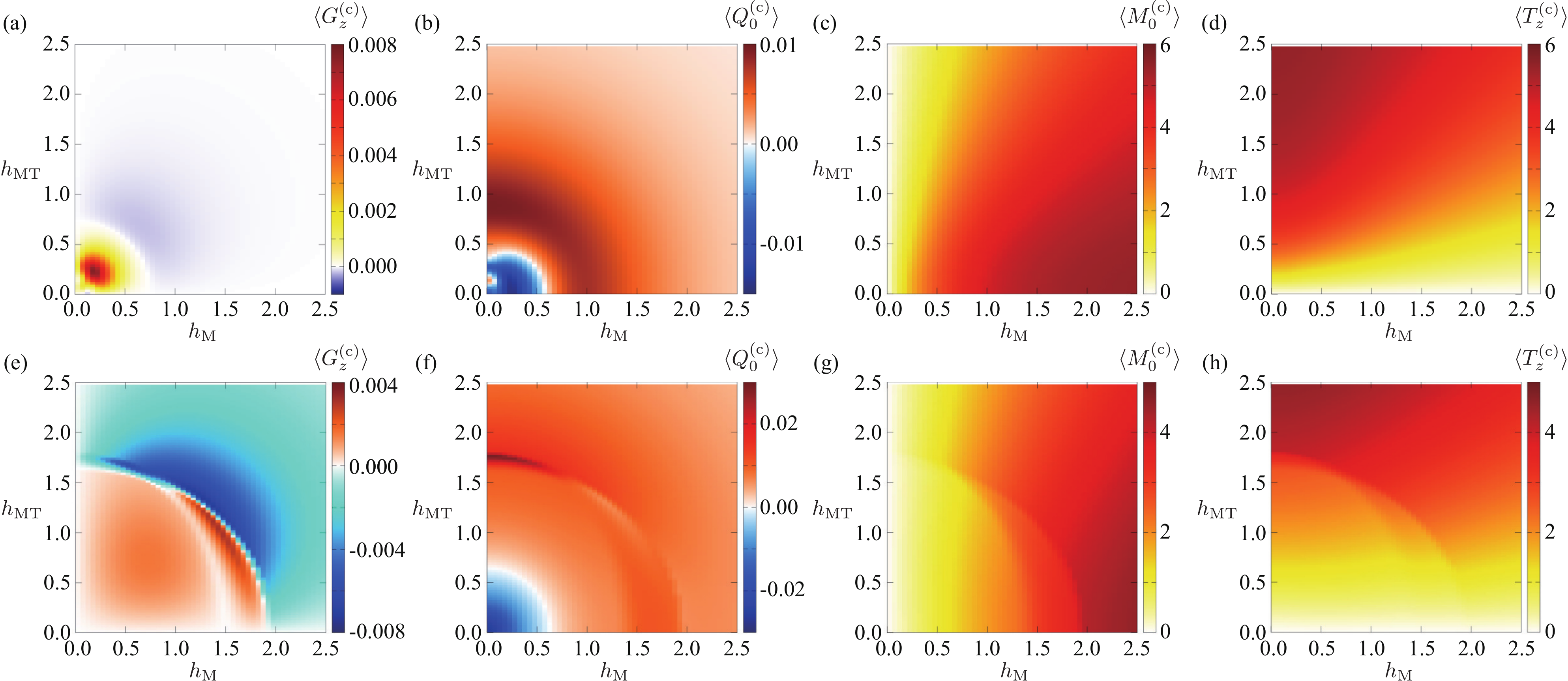} 
\caption{
\label{fig: Lattice_dat1}
Contour plots of (a,e) $\langle G_z^{\rm (c)} \rangle$, (b,f) $\langle Q_0^{\rm (c)} \rangle$, (c,g) $\langle M_0^{\rm (c)} \rangle$, and (d,h) $\langle T_z^{\rm (c)} \rangle$ in the plane of $h_{\rm M}$ and $h_{\rm MT}$ at (a)-(d) $\lambda=0.5$ and (e)-(h) $\lambda=5$ in the four-sublattice system. 
The other model parameters are chosen as $t=-1$, $t_p=0.7$, $t_{z}=0.2$, $t_{sp}=0.3$, $\Gamma=0.8$, and $\mu=0$. 
}
\end{center}
\end{figure*}

\section{Result}
\label{sec: Result}

We discuss the behavior of the ferro-axial moment against the model parameters. 
We present the results in the four-sublattice tetragonal system in Sec.~\ref{sec: Tetragonal case} and those in the single-site system in Sec.~\ref{sec: Atomic case}. 

\subsection{Four-sublattice tetragonal system}
\label{sec: Tetragonal case}

Figure~\ref{fig: Lattice_dat1} shows the contour plot of quantities in Eqs.~(\ref{eq:Q0c})-(\ref{eq:Tzc}) while changing $h_{\rm M}$ and $h_{\rm MT}$. 
The behaviors of $\langle G_z^{\rm (c)} \rangle$, $\langle Q_0^{\rm (c)} \rangle$, $\langle M_0^{\rm (c)} \rangle$, and $\langle T_z^{\rm (c)} \rangle$ are shown in Figs.~\ref{fig: Lattice_dat1}(a,e), \ref{fig: Lattice_dat1}(b,f), \ref{fig: Lattice_dat1}(c,g), and \ref{fig: Lattice_dat1}(d,h), respectively. 
The data in Figs.~\ref{fig: Lattice_dat1}(a)-\ref{fig: Lattice_dat1}(d) are calculated for the weak spin-orbit coupling $\lambda=0.5$ and those in Figs.~\ref{fig: Lattice_dat1}(e)-\ref{fig: Lattice_dat1}(h) are calculated for the large spin-orbit coupling $\lambda=5$. 
The other hopping parameters are taken at $t=-1$, $t_p=0.7$, $t_{z}=0.2$, $t_{sp}=0.3$, and $\Gamma=0.8$ and the chemical potential is set to be $\mu=0$. 
We take $N=1600^2$ here and hereafter. 
We also present the result while changing $\mu$ in Appendix~\ref{sec: Filling dependence of ferro-axial moment}.

The results in Figs.~\ref{fig: Lattice_dat1}(a) and \ref{fig: Lattice_dat1}(e) show that the ferro-axial moment corresponding to $\langle G_z^{\rm (c)} \rangle$ becomes nonzero under nonzero $h_{\rm M}$ and $h_{\rm MT}$; it vanishes for $h_{\rm M}=0$ or $h_{\rm MT}=0$. 
The sign change of $\langle G_z^{\rm (c)} \rangle$ around $\sqrt{h_{\rm M}^2+h_{\rm MT}^2} \sim 0.75$ for $\lambda=0.5$ and $\sqrt{h_{\rm M}^2+h_{\rm MT}^2} \sim 1.8$ for $\lambda=5$ is owing to the band crossing at $\mu=0$; the electron filling per site $n_{\rm e}=(1/4N)\langle \sum_{\bm{k}\gamma\alpha\sigma}c^{\dagger}_{\bm{k}\gamma\alpha\sigma}c^{}_{\bm{k}\gamma\alpha\sigma}\rangle$ roughly changes from 3 to 4, where $n_{\rm e}=8$ represents the full filling. 
In addition, one finds that $|\langle G_z^{\rm (c)} \rangle|$ tends to be larger when $h_{\rm M}$ is close to $h_{\rm MT}$, which indicates that both vortex spin configurations with $M_0$ and $T_z$ are important in inducing $\langle G_z^{\rm (c)} \rangle$, as discussed above. 

The data of $\langle Q_0^{\rm (c)} \rangle$, $\langle M_0^{\rm (c)} \rangle$, and $\langle T_z^{\rm (c)} \rangle$ are presented for reference in Figs.~\ref{fig: Lattice_dat1}(b)-(d) and \ref{fig: Lattice_dat1}(f)-(h). 
$\langle Q_0^{\rm (c)} \rangle$ in Figs.~\ref{fig: Lattice_dat1}(b) and \ref{fig: Lattice_dat1}(f) shows nonzero values for any $h_{\rm M}$ and $h_{\rm MT}$; $\langle Q_0^{\rm (c)} \rangle$ exists in the paramagnetic state for $h_{\rm M}=h_{\rm MT}=0$, since the electric monopole $Q_0^{\rm (c)} $ belongs to the totally symmetric representation under the $4/mmm 1'$ group. 
The emergence of $\langle Q_0^{\rm (c)} \rangle$ is owing to the local crystalline electric field that arises from the local inversion symmetry breaking at each sublattice, as shown in Fig.~\ref{fig: Lattice}(a). 
Meanwhile, $\langle M_0^{\rm (c)} \rangle$ ($\langle T_z^{\rm (c)} \rangle$) becomes nonzero unless $h_{\rm M}=0$ ($h_{\rm MT}=0$), which tends to be developed when increasing $h_{\rm M}$ ($h_{\rm MT}$).

\begin{figure}[htb!]
\begin{center}
\includegraphics[width=1.0 \hsize ]{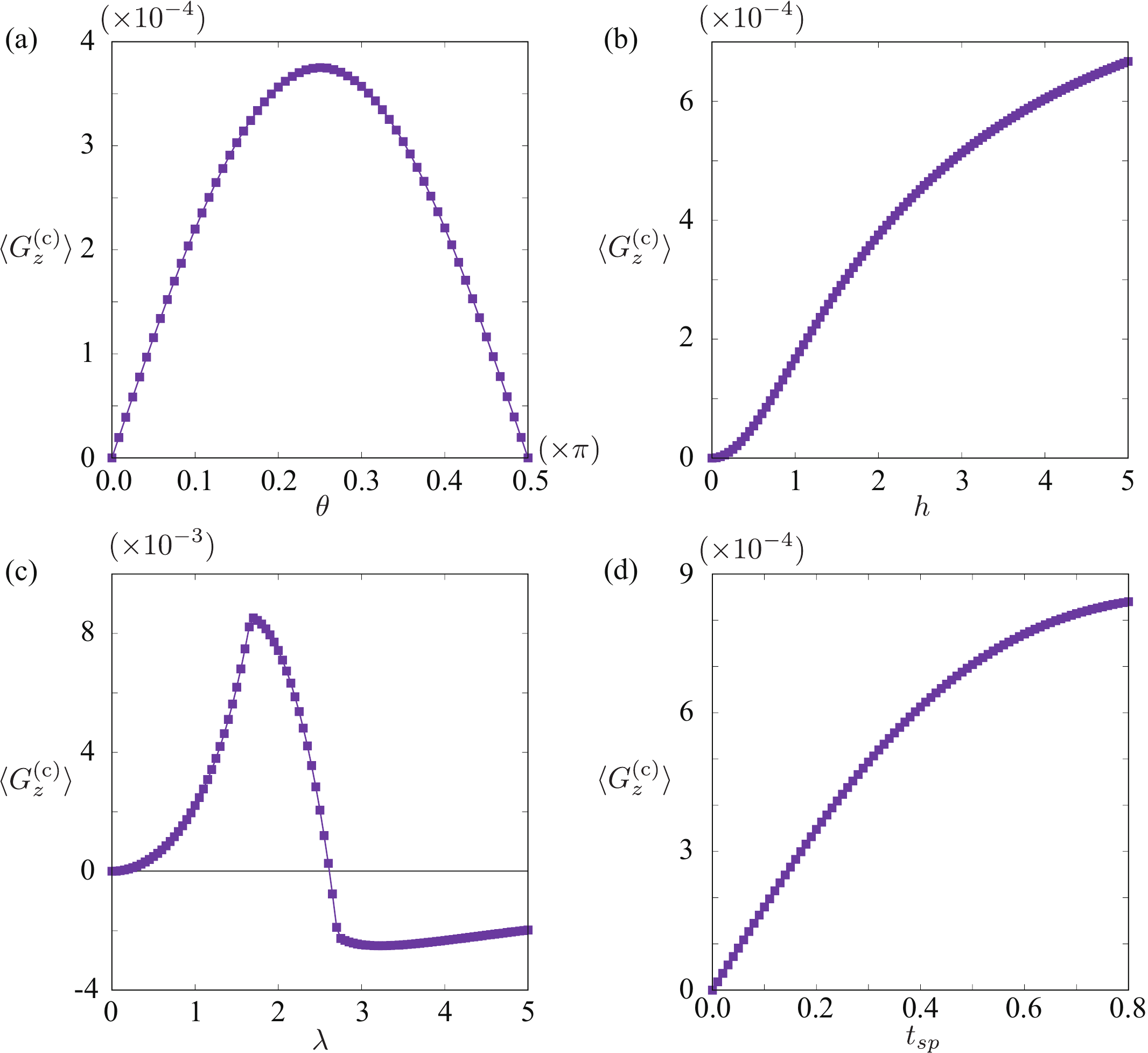} 
\caption{
\label{fig: Lattice_dat3}
The behavior of $\langle G_z^{\rm (c)} \rangle$ while changing (a) $\theta$, (b) $h$, (c) $\lambda$, and (d) $t_{sp}$ at $n_{\rm e}=0.2$. 
In (a), $h=2$, $t_{sp}=0.3$, and $\lambda=0.5$. 
In (b), $\theta=\pi/4$, $t_{sp}=0.3$, and $\lambda=0.5$. 
In (c), $h=\sqrt{2}$, $\theta=\pi/4$, and $t_{sp}=0.3$. 
In (d), $h=\sqrt{2}$, $\theta=\pi/4$, and $\lambda=0.5$. 
The other hopping parameters are common to those in Fig.~\ref{fig: Lattice_dat1}. 
}
\end{center}
\end{figure}

To extract the essential model parameters to induce $\langle G_z^{\rm (c)}\rangle$ under the vortex spin textures, we investigate the several model parameter dependences of $\langle G_z^{\rm (c)}\rangle$ by considering the low-filling $n_{\rm e}=0.2$ so that the Fermi surface is relatively simple. 
Figure~\ref{fig: Lattice_dat3}(a) shows the $\theta$ dependence of $\langle G_z^{\rm (c)}\rangle$ at $t=-1$, $t_p=0.7$, $t_{z}=0.2$, $t_{sp}=0.3$, $\Gamma=0.8$, $\lambda=0.5$, and $h=2$, where we introduce $\theta$ and $h$ instead of $h_{\rm M}$ and $h_{\rm MT}$ related as $h_{\rm M}=h \cos \theta$ and $h_{\rm MT}= h \sin \theta$. 
The data shows that $\langle G_z^{\rm (c)}\rangle$ takes nonzero values except for $\theta=0$ or $\theta=\pi/2$ and it seems to be almost symmetric against $\theta$, which is consistent with the result in Figs.~\ref{fig: Lattice_dat1}(a) and \ref{fig: Lattice_dat1}(e). 
Moreover, one finds that $\langle G_z^{\rm (c)}\rangle$ is developed while increasing $h$, as shown in the case of $\theta=\pi/4$ in Fig.~\ref{fig: Lattice_dat3}(b). 
These results clearly indicate that $\langle G_z^{\rm (c)}\rangle$ is induced by the vortex spin configuration with $\langle M_0^{\rm (c)}\rangle \neq 0$ and $\langle T_z^{\rm (c)}\rangle \neq 0$. 

In addition to $h_{\rm M}$ and $h_{\rm MT}$, we find that the spin-orbit coupling $\lambda$ and the hopping between $s$ and $p$ orbitals $t_{sp}$ are important for nonzero $\langle G_z^{\rm (c)}\rangle$, as shown in Figs.~\ref{fig: Lattice_dat3}(c) and \ref{fig: Lattice_dat3}(d), where $\langle G_z^{\rm (c)}\rangle$ vanishes for $\lambda=0$ or $t_{sp}=0$. 
The mean-field parameters are taken at $h=\sqrt{2}$ and $\theta=\pi/4$ ($h_{\rm M}=h_{\rm MT}=2$). 
The necessity of $t_{sp}$ is reasonable since the local electric dipole degree of freedom is described by the $s$-$p$ hybridization as found in Eq.~(\ref{eq: localQ}); the $s$-orbital Hilbert space is completely decoupled from the $p$-orbital one when $t_{sp}=0$. 
Indeed, $\langle Q_0^{\rm (c)}\rangle$ also vanishes for $t_{sp}=0$. 
On the other hand, the necessity of $\lambda$ seems to be rather nontrivial, since the local electric polarization $\bm{Q}_{\gamma}$ does not have a spin dependence. 
In contrast to $t_{sp}$, $\langle Q_0^{\rm (c)}\rangle$ becomes nonzero for $\lambda=0$. 
In other words, the result implies that $t_{sp}$ plays a role in inducing the local electric polarization $\bm{Q}_\eta$ and $\lambda$ plays a role in tilting $\bm{Q}_\eta$ so that $\langle G_z^{\rm (c)}\rangle$ becomes nonzero.

\subsection{Single-site system}
\label{sec: Atomic case}

\begin{figure}[t!]
\begin{center}
\includegraphics[width=1.0 \hsize ]{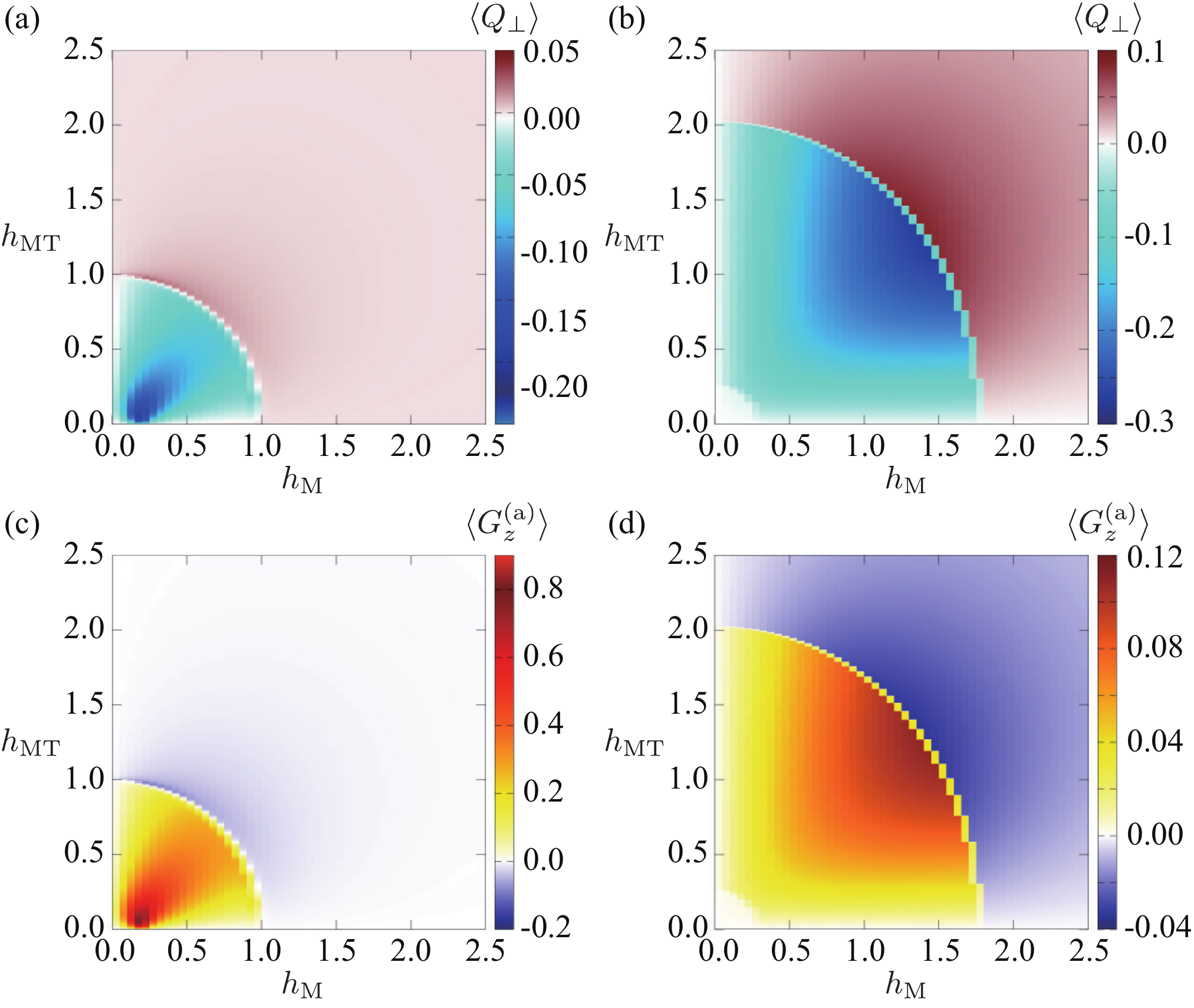} 
\caption{
\label{fig: atomic_dat}
Contour plots of (a,b) $\langle Q_{\perp} \rangle$ and (c,d) $\langle G_z^{\rm (a)} \rangle$ in the plane of $h_{\rm M}$ and $h_{\rm MT}$ at (a,c) $\lambda=0.5$ and (b,d) $\lambda=5$ in the single-site system. 
The other model parameters are chosen at $V_{sp}=1$ and $\mu=0$.
}
\end{center}
\end{figure}

To further discuss the minimum essence to induce $\langle G_z^{\rm (c)}\rangle$ under the vortex spin configuration, we consider the single-site system by setting $t=t_p=t_z=t_{sp}=0$; we focus on the sublattice A. 
In addition, to take into account the effect of the $s$-$p$ hybridization that is necessary for obtaining nonzero $\langle G_z^{\rm (c)}\rangle$, we introduce the local $s$-$p$ hybridization, whose Hamiltonian is given by 
\begin{align}
\mathcal{H}^{s-p}=V_{sp}\sum_{\sigma} (c^{\dagger}_{{\rm A}s\sigma}c^{}_{{\rm A}p_x\sigma}+c^{\dagger}_{{\rm A}s\sigma}c^{}_{{\rm A}p_y\sigma})+{\rm h.c.}, 
\end{align}
where we drop off the irrelevant subscript $\bm{k}$. 
In the end, the independent model parameters in the single-site system are $\lambda$, $V_{sp}$, $h_{\rm M}$, and $h_{\rm MT}$. 

Figures~\ref{fig: atomic_dat}(a) and \ref{fig: atomic_dat}(b) show the results at $\lambda=0.5$ and $\lambda=5$, respectively, for $V_{sp}=1$ and $\mu=0$. 
We present the contour plot of $\langle Q_{\perp} \rangle = (\bm{e}_{\rm A} \times \langle \bm{Q}_{\rm A} \rangle)_z$ in the plane of $h_{\rm M}$ and $h_{\rm MT}$, which corresponds to the A sublattice component of $\langle G_z^{\rm (c)}\rangle$ in Eq.~(\ref{eq:Gzc}); $\langle Q_{\perp} \rangle$ means the electric polarization perpendicular to the position vector. 
The overall qualitative behaviors in Figs.~\ref{fig: atomic_dat}(a) and \ref{fig: atomic_dat}(b) are similar to those in Figs.~\ref{fig: Lattice_dat1}(a) and \ref{fig: Lattice_dat1}(e), respectively; $\langle Q_{\perp} \rangle$ becomes nonzero for $h_{\rm M} \neq 0$ and $h_{\rm MT} \neq 0$. 
In other words, the single-site system rather than the multi-sublattice one is enough to describe the emergence of the ferro-axial moment under the magnetic ordering. 
This result indicates that there are four important parameters to induce nonzero $\langle Q_{\perp} \rangle$; $\lambda$, $V_{sp}$, $h_{\rm M}$, and $h_{\rm MT}$. 
When setting any of $\lambda$, $V_{sp}$, $h_{\rm M}$, and $h_{\rm MT}$ to be zero, $\langle Q_{\perp} \rangle$ vanishes. 

To further confirm the necessity of four model parameters to obtain $\langle Q_{\perp} \rangle$, we expand it as a polynomial form of products of the Hamiltonian matrix based on the procedure in Refs.~\cite{Hayami_PhysRevB.102.144441, Oiwa_doi:10.7566/JPSJ.91.014701}. 
As a result, we obtain the lowest contribution to $\langle Q_{\perp} \rangle$ as the 5th order, which is proportional to $h_{\rm M} h_{\rm MT}  V_{sp} \lambda^2 $. 
Moreover, we find that the expansion includes the same factor $h_{\rm M} h_{\rm MT}  V_{sp} \lambda^2 $ in the higher-order contribution, at least, up to the 10th order. 
Thus, the essential model parameters to cause nonzero $\langle Q_{\perp} \rangle$ are given by $h_{\rm M} h_{\rm MT}  V_{sp} \lambda^2 \sim h^2 V_{sp} \lambda^2 \sin 2\theta$. 
Indeed, such model-parameter dependences are consistent with the behavior of $\langle G_z^{\rm (c)} \rangle$ in the region where the Fermi surface geometry is not important in Figs.~\ref{fig: Lattice_dat3}(a)-\ref{fig: Lattice_dat3}(d).

Notably, we also find that another atomic-scale quantity $\langle G^{\rm (a)}_z \rangle$, which is different from the electric polarization $\langle Q_{\perp} \rangle$, is induced when the ferro-axial moment appears. 
This quantity is defined as 
\begin{align}
\label{eq: G_atomic}
G^{\rm (a)}_z = (\bm{l} \times \bm{\sigma})_z, 
\end{align}
where $\bm{l}$ is the orbital angular momentum operator for the $p$ orbital~\cite{hayami2021electric}. 
The expression of $G^{\rm (a)}_z$ in Eq.~(\ref{eq: G_atomic}) is derived based on a complete multipole basis set for the atomic-scale wave function~\cite{kusunose2020complete}.
As both $\bm{l}$ and $\bm{\sigma}$ correspond to the axial-vector quantities with time-reversal odd, their cross product results in the axial-vector quantity with time-reversal even like the electric toroidal dipole, which is similar to $G^{\rm (c)}_z$.
On the other hand, in contrast to $\langle G^{\rm (c)}_z \rangle$ in Eq.~(\ref{eq:Gzc}), $G^{\rm (a)}_z$ does not depend on the choice of the origin in the unit cell, as it is an atomic-scale quantity.
Although it is difficult to directly observe $G^{\rm (a)}_z$ because it is not a conjugate quantity to the electromagnetic field, it becomes a source of off-diagonal responses including the spin-current generation~\cite{hayami2021electric, Roy_PhysRevMaterials.6.045004} and antisymmetric thermopolarization~\cite{Nasu_PhysRevB.105.245125}.
Figures~\ref{fig: atomic_dat}(c) and \ref{fig: atomic_dat}(d) show the result of $\langle G^{\rm (a)}_z \rangle$, whose behavior is similar to that of $\langle Q_{\perp} \rangle$ in Figs.~\ref{fig: atomic_dat}(a) and \ref{fig: atomic_dat}(b).

\begin{table}[htb!]
\centering
\begingroup
\renewcommand{\arraystretch}{2.0}
\caption{
Active multipoles in the $s$-$p$ hybridized orbital system with the spin degree of freedom; $j$ represents the total orbital angular momentum. 
$X_0$, $X_{1m}$, $X_{2m}$, and $X_{3m}$ for $X=Q$, $M$, $T$, and $G$ stand for the monopole, dipole, quadrupole, and octupole, respectively. 
\label{table: atomic}}
\begin{tabular}{ccccccccccccc}\hline \hline
 & $\displaystyle j=\frac{1}{2}$ $(s)$  & $\displaystyle j=\frac{1}{2}$ $(p)$ & $\displaystyle j=\frac{3}{2}$ $(p)$   \\
\hline
$\displaystyle j=\frac{1}{2}$ $(s)$ & $Q_0 \oplus M_{1m}$ & $G_0  \oplus Q_{1m} $ &  $Q_{1m}  \oplus G_{2m} $ \\ 
$\displaystyle j=\frac{1}{2}$ $(p)$ & $M_0  \oplus T_{1m} $ & $Q_0 \oplus M_{1m}$ & $G_{1m}  \oplus Q_{2m} $ \\
$\displaystyle j=\frac{3}{2}$ $(p)$ & $T_{1m}  \oplus M_{2m} $  & $M_{1m}  \oplus T_{2m} $ &  $Q_{0}  \oplus Q_{2m} \oplus M_{1m}  \oplus M_{3m} $
 \\
\hline\hline
\end{tabular}
\endgroup
\end{table}

It is noted that such an atomic-scale $G^{\rm (a)}_z$ is the only electronic degree of freedom in the $s$-$p$ hybridized orbital system with the spin degree of freedom. 
In the single-site model, as a physical Hilbert space is spanned by the eight basis wave functions with four orbital and two spin degrees of freedom, there are $8\times 8=64$ electronic degrees of freedom in the model Hamiltonian. 
We show the multipole degrees of freedom corresponding to these 64 electronic ones in Table~\ref{table: atomic}, where $X_0$, $X_{1m}$, $X_{2m}$, and $X_{3m}$ for $X=Q$, $M$, $T$, and $G$ stand for the monopole, dipole, quadrupole, and octupole, respectively. 
In the table, the Hilbert space where the multipoles become active is presented. 
For example, in the Hilbert space spanned by the $s$ orbital with $j=1/2$ and the $p$ orbital with $j=1/2$, the electric toroidal monopole $G_0$, electric dipole $Q_{1m}$, magnetic monopole $M_0$, and magnetic toroidal dipole $T_{1m}$ become active. 
Among the active multipoles, only the electric toroidal dipole $G_{1m}$ (or $\bm{G}$) corresponds to the time-reversal-even axial-vector degree of freedom in the single-site system. 
We also present the classification of the active multipoles in Table~\ref{table: atomic} under the $D_{4\rm h}$ group in Table~\ref{tab: irrep_D4h}, where $(X_x, X_y, X_z)$, $(X_{u}, X_{v}, X_{yz}, X_{zx}, X_{xy})$, and $(X_{xyz}, X_x^\alpha, X_y^\alpha, X_z^\alpha, X_x^\beta, X_y^\beta, X_z^\beta)$ stands for the dipole, quadrupole, and octupole components, respectively; $G_z$ belonging to the irreducible representation $A^+_{2g}$ is independent of the other multipoles. 
It is noted that the atomic quantity $Q_{\perp}$ belongs to the irreducible representation $E^+_u$ rather than $A^+_{2g}$ under the $D_{4\rm h}$ group, although the cluster structure of $Q_{\perp}$, i.e., $G^{\rm (c)}_{z}$, belongs to the irreducible representation $A^+_{2g}$.

\begin{table}
\caption{
Classification of the active multipoles in Table~\ref{table: atomic} under the tetragonal point group $D_{\rm 4h}$. 
The superscript $+$ in terms of the time-reversal parity is supposed for electric (E) and electric toroidal (ET) multipoles, while $-$ is supposed for magnetic (M) and magnetic toroidal (MT) multipoles. 
}
\label{tab: irrep_D4h}
\centering
\begin{tabular}{c|llllc} \hline\hline
$D_{\rm 4h}$ &
E & ET & M & MT 
\\ \hline
$A_{1g}$ &
$Q_{0}$, $Q_{u}$ & --- & --- &  $T_{u}$ &
 \\
$A_{2g}$ &
--- & $G_{z}$ & $M_{z}$, $M_{z}^{\alpha}$ & --- &
 \\
$B_{1g}$ &
$Q_{v}$ & --- & $M_{xyz}$ & $T_{v}$ &
 \\
$B_{2g}$ &
$Q_{xy}$ & --- & $M_{z}^{\beta}$ & $T_{xy}$ &
 \\
 $E_{g}$ &
$Q_{yz}$ & $G_{x}$  & $M_{x}$, $M_{x}^{\alpha}$, $M_{x}^{\beta}$ &  $T_{yz}$ &
 \\
& 
$Q_{zx}$ & $G_{y}$ & $M_{y}$, $M_{y}^{\alpha}$, $M_{y}^{\beta}$ & $T_{zx}$  &
 \\
 \hline
$A_{1u}$ &
--- & $G_{0}$, $G_{u}$ & $M_{0}$, $M_{u}$ & --- &
 \\
$A_{2u}$ &
$Q_{z}$ & --- & ---  & $T_{z}$ &
 \\
$B_{1u}$ &
--- & $G_{v}$ & $M_{v}$ & --- &
 \\
$B_{2u}$ &
--- & $G_{xy}$ & $M_{xy}$ & --- &
 \\
$E_{u}$ &
$Q_{x}$ &  $G_{yz}$ & $M_{yz}$  & $T_{x}$ &
 \\
& 
$Q_{y}$ &  $G_{zx}$ & $M_{zx}$  & $T_{y}$ &
 \\
\hline\hline
\end{tabular}
\end{table}

By performing a similar procedure to extracting the model-parameter dependences of $\langle Q_{\perp} \rangle$, we obtain the essential model parameters for $G^{\rm (a)}_z$, which are given in the form of $h_{\rm M}h_{\rm MT} V^2_{sp} \lambda$ in the lowest order. 
By comparing the model-parameter dependences between $\langle Q_{\perp} \rangle$ and $G^{\rm (a)}_z$, one obtains the relation as follows: 
\begin{align}
\label{eq: propto}
\frac{\langle Q_{\perp} \rangle}{\langle G^{\rm (a)}_z\rangle}
=-\frac{\lambda}{2V_{sp}}. 
\end{align}
The relation holds, at least, up to the 10th order in the expansion, and is satisfied in the numerical results as shown in Fig.~\ref{fig: atomic_dat}. 
Thus, the atomic-scale electric toroidal dipole (ferro-axial moment) is closely related to the perpendicular component of the electric polarization, the latter of which corresponds to the conventional expression of the electric toroidal dipole defined by the vortex texture of the electric polarization. 

In addition, let us remark on the minimum orbital degree of freedom to induce the ferro-axial moment. 
As we consider the vortex texture of $\bm{Q}_\eta$ in the $xy$ plane, the $s$, $p_x$, and $p_y$ orbitals are the necessary ingredients. 
Meanwhile, we find that the $p_z$ orbital degree of freedom is also important for nonzero $\langle Q_{\perp} \rangle$ and $G^{\rm (a)}_z$. 
This is because the in-plane components of $\bm{l}$ included in $G^{\rm (a)}_z$ [Eq.~(\ref{eq: G_atomic})] are characterized by the off-diagonal matrix elements in Hilbert space between $(p_x, p_y)$ and $p_z$ orbitals. 
Thus, $G^{\rm (a)}_z$ is no longer activated once the $p_z$ orbital degree of freedom is neglected. 
As $\langle Q_{\perp} \rangle$ is proportional to $G^{\rm (a)}_z$ in Eq.~(\ref{eq: propto}), $\langle Q_{\perp} \rangle$ also vanishes in such a situation. 

\section{Relevance with skyrmion}
\label{sec: Relevance with skyrmion}

\begin{figure}[htb!]
\begin{center}
\includegraphics[width=1.0 \hsize ]{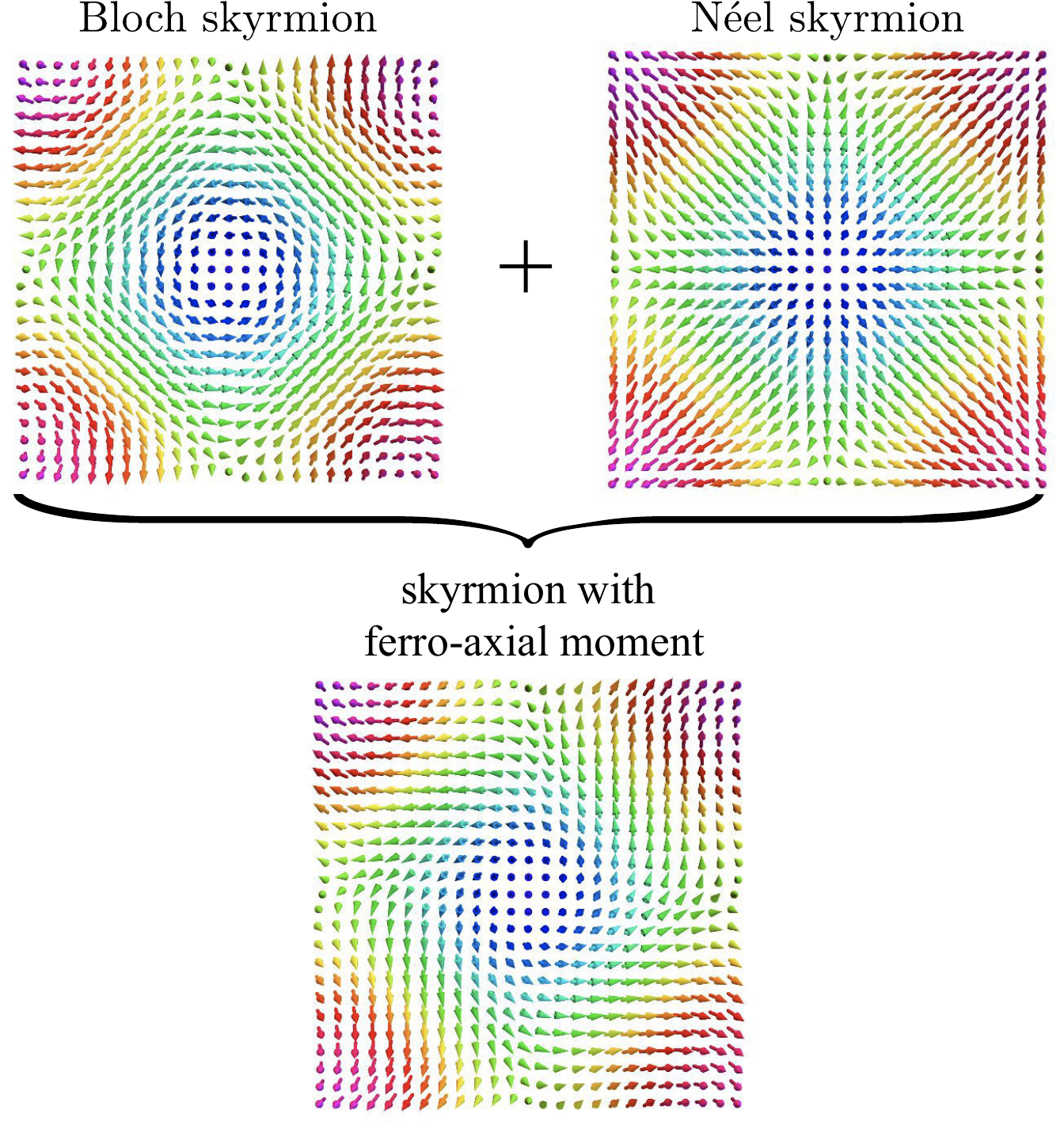} 
\caption{
\label{fig: SkX}
The ferro-axial nature ($G_z$) under the superposition (lower panel) of the Bloch skyrmion with $T_z$ (upper-left panel) and N\'eel skyrmion with $M_0$ (upper-right panel). 
The arrow represents the spins, where the blue, green, and red colors stand for down, zero, and up spins, respectively. 
}
\end{center}
\end{figure}

So far, we have focused on the ferro-axial moment under the vortex with the coplanar spin configuration, as shown in Fig.~\ref{fig: ponti}(b). 
In this section, we argue that a magnetic skyrmion with a topologically-nontrivial noncoplanar spin texture is another candidate to exhibit the ferro-axial nature. 
We show the schematic pictures of two types of skyrmions in Fig.~\ref{fig: SkX}: One is the Bloch skyrmion in the upper-left panel and the other is the N\'eel skyrmion in the upper-right panel. 
They are distinguished by the helicity around the skyrmion core located at the centering down spin in blue. 
From the multipole description, the Bloch skyrmion accompanies with $T_z$, while the N\'eel skyrmion accompanies with $M_0$~\cite{Gobel_PhysRevB.99.060406, Hayami_PhysRevB.105.104428,Bhowal_PhysRevLett.128.227204}.
Thus, by considering a superposition of the Bloch skyrmion and N\'eel skyrmion, one can induce the ferro-axial moment under the magnetic ordering, as shown in Fig.~\ref{fig: SkX}. 

Depending on the presence/absence of the spatial inversion symmetry, we present two scenarios to realize the skyrmion with the ferro-axial moment. 
One is based on the Dzyaloshinskii-Moriya (DM) interaction in the noncentrosymmetric system. 
For example, one can consider the competition between the chiral-type DM interaction and the polar-type DM interaction. 
As the former (latter) tends to stabilize the Bloch (N\'eel) skyrmion, the noncentrosymmetric system with chiral- and polar-type DM interactions, such as the point groups $C_6$ and $C_4$, is a prototypical system~\cite{Banerjee_PhysRevX.4.031045,oh2014effects,Rowland_PhysRevB.93.020404,Garlow_PhysRevLett.122.237201,peng2021tunable}. 
Moreover, even in the purely polar (or chiral) system, the skyrmion spin texture with the ferro-axial moment can be realized when additionally considering other anisotropic exchange interactions that are allowed from the lattice symmetry, e.g., the bond-dependent exchange interaction~\cite{Hayami_PhysRevLett.121.137202}. 
The van der Waals magnets, such as Fe$_3$GeTe$_2$, are the candidates belonging to this category~\cite{peng2021tunable}. 

The other is based on the competing exchange interaction in the centrosymmetric system where the DM interaction does not play an important role. 
The typical mechanisms are represented by the short-range exchange interactions in the localized spin system~\cite{Okubo_PhysRevLett.108.017206,leonov2015multiply, Lin_PhysRevB.93.064430, Hayami_PhysRevB.93.184413, Hayami_PhysRevB.94.174420} and the long-range exchange interaction that arises from the itinerant nature of electrons in the itinerant electron system~\cite{Ozawa_PhysRevLett.118.147205, Hayami_PhysRevB.95.224424, Hayami_PhysRevB.99.094420, Wang_PhysRevLett.124.207201,wang2021skyrmion,hayami2021topological,Hayami_10.1088/1367-2630/ac3683,Eto_PhysRevLett.129.017201}. 
In this case, the helicity of the skyrmion is determined by magnetic anisotropy~\cite{amoroso2020spontaneous,yambe2021skyrmion, Hayami_PhysRevB.103.024439, Hayami_PhysRevB.103.054422, Wang_PhysRevB.103.104408, amoroso2021tuning} and the dipolar interaction~\cite{Utesov_PhysRevB.103.064414,Utesov_PhysRevB.105.054435}. 
Especially, it was revealed that the skyrmion corresponding to the lower panel of Fig.~\ref{fig: SkX} can be realized by considering a frustration arising from the momentum-resolved interaction under the $D_{4\rm h}$ point group~\cite{Hayami_doi:10.7566/JPSJ.89.103702,hayami2022multiple} and the $C_{4\rm h}$ point group~\cite{Hayami_PhysRevB.105.104428}, or by taking into account the staggered DM interaction under the $D_{6\rm h}$ point group~\cite{Hayami_PhysRevB.105.184426}. 
The SkX-hosting centrosymmetric materials, such as Gd$_2$PdSi$_3$~\cite{kurumaji2019skyrmion,Kumar_PhysRevB.101.144440,spachmann2021magnetoelastic}, Gd$_3$Ru$_4$Al$_{12}$~\cite{hirschberger2019skyrmion,Hirschberger_10.1088/1367-2630/abdef9}, GdRu$_2$Si$_2$~\cite{khanh2020nanometric,Yasui2020imaging,khanh2022zoology}, and EuAl$_4$~\cite{Shang_PhysRevB.103.L020405,kaneko2021charge,Zhu_PhysRevB.105.014423,takagi2022square}, are candidates in this category.

\section{Summary}
\label{sec: Summary}

To summarize, we have investigated an essence to induce the ferro-axial nature under magnetic orderings. 
We show that the superposition of the vortices with different helicities so as to have magnetic monopole $M_0$ and magnetic toroidal dipole $T_z$ naturally leads to the ferro-axial moment $G_z$. 
We also present all the magnetic point groups to possess $M_0$, $T_z$, and $G_z$ from the symmetry viewpoint. 
Then, by considering a minimum tetragonal model, we demonstrate that the vortex spin textures exhibit $G_z$ in a four-sublattice cluster. 
We clarify that the interplay among the magnetic order parameters, the atomic spin-orbit coupling, and the hybridization between orbitals with different parity is an essence to induce $G_z$. 
Furthermore, we show that the ferro-axial moment becomes nonzero even in the single-site system, where it is described by the vector product of the orbital and spin angular momentum operators. 
Finally, we discuss a possible realization of the magnetic-order-driven ferro-axial moment by exemplifying the skyrmion. 

\begin{table*}[htb!]
\centering
\caption{
Candidate materials to possess the ferro-axial moment under the magnetic orderings. 
\label{table: Materials}}
\begin{tabular}{cl}\hline \hline
MPG & Materials\\
\hline 
$6/m'$ & U$_{14}$Au$_{51}$~\cite{brown1997structure} \\
$6$ & ScMnO$_3$~\cite{Munoz_PhysRevB.62.9498}, Yb$_{0.42}$Sc$_{0.58}$FeO$_3$~\cite{Tang_PhysRevB.103.174102}, BaCoSiO$_4$~\cite{ding2021field} \\
$\bar{6}'$ & Cu$_{0.82}$Mn$_{1.18}$As~\cite{Karigerasi_PhysRevMaterials.3.111402}, Tb$_{14}$Ag$_{51}$~\cite{pomjakushin2022revisiting} \\ \hline
$\bar{3}'$ & MgMnO$_3$~\cite{Haraguchi_PhysRevMaterials.3.124406}, Yb$_3$Pt$_4$~\cite{Janssen_PhysRevB.81.064401}  \\
$3$ & Cu$_2$OSeO$_3$~\cite{Bos_PhysRevB.78.094416}, Mn$_2$FeMoO$_6$~\cite{li2014magnetic} \\ \hline
$4/m'$ & (K,Rb)$_y$Fe$_{2-x}$Se$_2$~\cite{pomjakushin2011room}, TlFe$_{1.6}$Se$_2$~\cite{May_PhysRevLett.109.077003}, K$_{0.8}$Fe$_{1.8}$Se$_2$~\cite{wei2011novel}, NdB$_4$~\cite{yamauchi2017magnetic} \\
$4$ & Ce$_5$TeO$_8$~\cite{podchezertsev2021influence} \\
$\bar{4}'$ & CsCoF$_4$~\cite{lacorre1991ordered}  \\ \hline
$2/m'$ & LiFePO$_4$~\cite{Toft_PhysRevB.92.024404}, (Co, Fe)$_4$Nb$_2$O$_9$~\cite{Khanh_PhysRevB.93.075117,Khanh_PhysRevB.96.094434,Deng_PhysRevB.97.085154,Yanagi_PhysRevB.97.020404,Ding_PhysRevB.102.174443}, ErGe$_3$~\cite{schobinger1996magnetic}, CaMnGe~\cite{welter1996neutron}, KFeSe$_2$~\cite{bronger1987antiferromagnetic}, Fe$_2$Co$_2$Nb$_2$O$_9$~\cite{maignan2021fe}
 \\
$2$ & LiFeP$_2$O$_7$~\cite{rousse2002neutron}, SrMn(VO$_4$)(OH)~\cite{Sanjeewa_PhysRevB.93.224407}, DyCrWO$_6$~\cite{ghara2018synthesis}, Ba$_3$MnSb$_2$O$_9$~\cite{doi2004structural}, HoNiO$_3$~\cite{Fernandes_PhysRevB.64.144417} \\ 
$m'$ & MnTiO$_3$~\cite{Arevalo_PhysRevB.88.104416}, ScFeO$_3$~\cite{li2012polar}, GaFeO$_3$~\cite{niu2017room}, Ce$_2$PdGe$_3$~\cite{cedervall2016low}, Mn$_3$O$_4$~\cite{boucher1971proprietes}  \\
$\bar{1}'$ & CaMnGe$_2$O$_6$~\cite{redhammer2008magnetic}, MnPSe$_3$~\cite{wiedenmann1981neutron,Calder_PhysRevB.103.024414}, BaNi$_2$P$_2$O$_8$~\cite{regnault1980magnetic}, YbMn$_2$Sb$_2$~\cite{morozkin2006synthesis}, NaCrSi$_2$O$_6$~\cite{Nenert_PhysRevB.81.184408}, CaMn$_2$Sb$_2$~\cite{bridges2009magnetic}	\\
$1$ & CuB$_2$O$_4$~\cite{Boehm_PhysRevB.68.024405}
 \\
\hline\hline
\end{tabular}
\end{table*}

We list the candidate materials to exhibit the ferro-axial moment under the magnetic orderings in Table~\ref{table: Materials} in accordance with MAGNDATA, the magnetic structures database~\cite{gallego2016magndata,gallego2016magndata2}.
The materials hosting the vortex spin configurations in the hexagonal crystal structures, U$_{14}$Au$_{51}$~\cite{brown1997structure}, ScMnO$_3$~\cite{Munoz_PhysRevB.62.9498}, BaCoSiO$_4$~\cite{ding2021field}, Cu$_{0.82}$Mn$_{1.18}$As~\cite{Karigerasi_PhysRevMaterials.3.111402}, and Tb$_{14}$Ag$_{51}$~\cite{pomjakushin2022revisiting}, the trigonal crystal structure, Yb$_3$Pt$_4$~\cite{Janssen_PhysRevB.81.064401}, and the tetragonal crystal structure, NdB$_4$~\cite{yamauchi2017magnetic}, might be prototypes to possess the ferro-axial moment. 
In addition, the ferro-axial moment can be induced even in the collinear spin configurations once the magnetic monopole and magnetic toroidal dipole degrees of freedom are activated under the magnetic orderings. 
Table~\ref{table: Materials} includes such magnetic materials for future exploration. 
In these materials, one can expect physical phenomena characteristic of ferro-axial moment, such as the spin-current generation~\cite{hayami2021electric} and antisymmetric thermopolarization~\cite{Nasu_PhysRevB.105.245125}, which will stimulate further exploration of the ferro-axial-related physical phenomena.

\appendix
\section{Ferro-axial moment under anti-vortex spin textures}
\label{sec: Ferro-axial moment under anti-vortex spin textures}

\begin{figure}[htb!]
\begin{center}
\includegraphics[width=1.0 \hsize ]{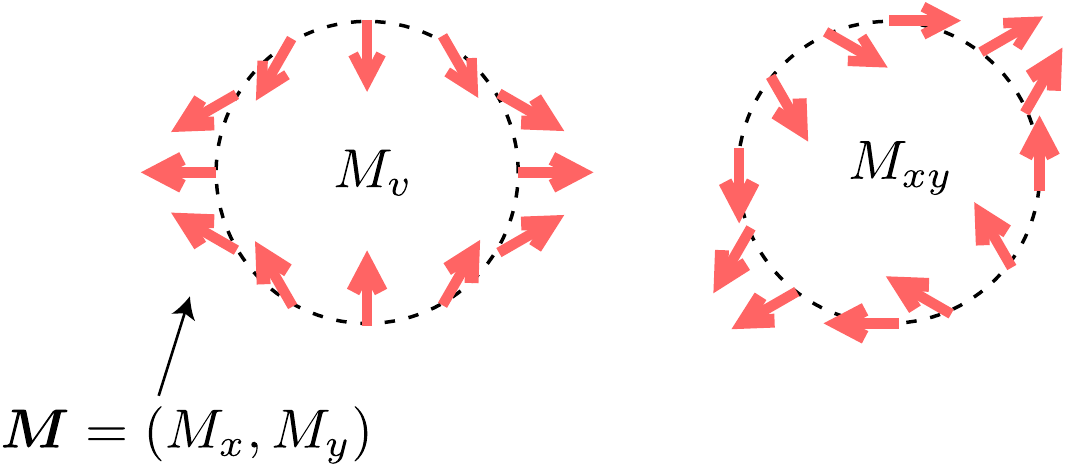} 
\caption{
\label{fig: antivortex}
Anti-vortex configurations of the magnetic dipoles $\bm{M}=(M_x, M_y, 0)$. 
The left and right panels show the different components of the magnetic quadrupoles, $M_{v}$ and $M_{xy}$, respectively. 
}
\end{center}
\end{figure}

\begin{table}[t!]
\centering
\caption{
\label{table: MPG2}
Reduction from the tetragonal gray point group (GPG) to the subgroups when nonzero $M_{v}$, $T_{xy}$, and $G_z$ appear. 
The other active multipoles, $M_z$, $G_0$, $Q_z$, and $T_0$, are also presented by $\checkmark$. 
}
\begin{tabular}{ccccccccccccc}\hline \hline
GPG & $M_v, M_{xy}, G_z$  & $M_z$ & $G_0$ & $Q_z$ & $T_0$\\
\hline 
$4/mmm 1'$, $4/m 1'$  & $4'/m'$ & -- & -- & -- & -- \\
$422 1'$, $4mm 1'$, $4 1'$ & $4'$ & -- & $\checkmark$ & $\checkmark$ & -- \\
$\bar{4}2m 1'$, $\bar{4}1'$ & $\bar{4}$ & $\checkmark$ & -- & -- & $\checkmark$  
 \\
\hline\hline
\end{tabular}
\end{table}

In this Appendix, we show that a superposition of the anti-vortex with two types of magnetic quadrupoles $M_v$ and $M_{xy}$ also leads to the ferro-axial moment. 
Figure~\ref{fig: antivortex} shows the schematic vortex spin configurations with $M_v$ and $M_{xy}$, whose vorticity is opposite to that with $M_0$ and $T_z$. 
By using $\bm{r}=(x,y,z)$, the expressions of $M_v$ and $M_{xy}$ are represented in the rank-2 symmetric form as $x M_x -y M_y$ and $x M_y + y M_x$, respectively~\cite{Prosandeev_PhysRevB.77.060101}. 
Similar to the situation in Fig.~\ref{fig: ponti2}(a), the spin configuration characterized by a superposition of two anti-vortices induces the ferro-axial moment. 
Such a superposition is especially expected in the tetragonal system or its subgroups, where the irreducible representations of $M_v$ and $M_{xy}$ are one-dimensional. 
We summarize the symmetry reduction from the tetragonal point group to the subgroups with nonzero $M_v$, $M_{xy}$, and $G_z$ in Table~\ref{table: MPG2}. 
One of the candidate materials belonging to the magnetic point groups in Table~\ref{table: MPG2} is K$X$O$_4$ ($X=$ Os, Ru)~\cite{marjerrison2016structure, Hayami_PhysRevB.97.024414,injac2019structural, Yamaura_PhysRevB.99.155113}. 

\section{Filling dependence of ferro-axial moment}
\label{sec: Filling dependence of ferro-axial moment}

\begin{figure*}[htb!]
\begin{center}
\includegraphics[width=1.0 \hsize ]{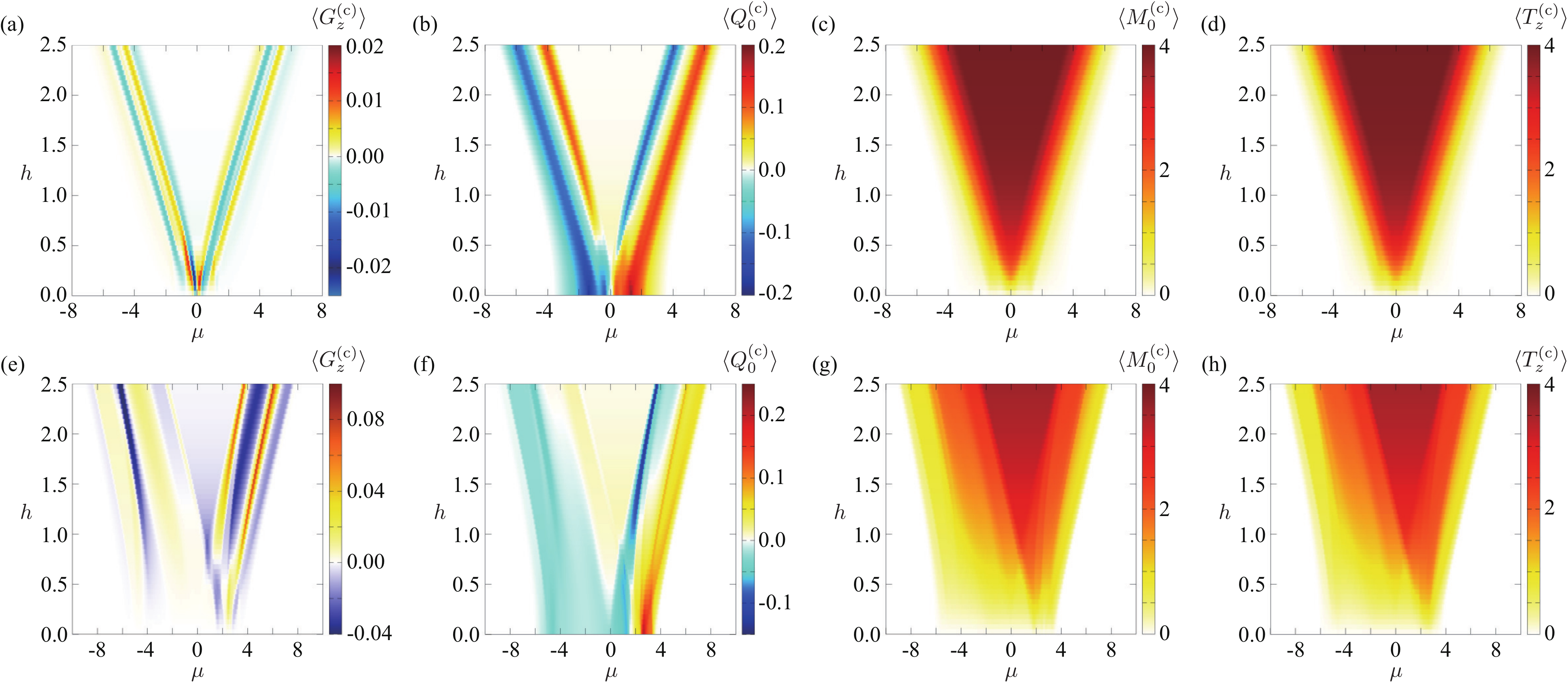} 
\caption{
\label{fig: Lattice_dat2}
Contour plots of (a,e) $\langle G_z^{\rm (c)} \rangle$, (b,f) $\langle Q_0^{\rm (c)} \rangle$, (c,g) $\langle M_0^{\rm (c)} \rangle$, and (d,h) $\langle T_z^{\rm (c)} \rangle$ in the plane of $\mu$ and $h$ at (a)-(d) $\lambda=0.5$ and (e)-(h) $\lambda=5$. 
The other model parameters are chosen as $t=-1$, $t_p=0.7$, $t_{z}=0.2$, $t_{sp}=0.3$, $\Gamma=0.8$, and $\theta=\pi/4$. 
}
\end{center}
\end{figure*}

We show the contour plot of $\langle G_z^{\rm (c)} \rangle$, $\langle Q_0^{\rm (c)} \rangle$, $\langle M_0^{\rm (c)} \rangle$, and $\langle T_z^{\rm (c)} \rangle$ for $\lambda=0.5$ and $5$ in Fig.~\ref{fig: Lattice_dat2} while changing the chemical potential $\mu$ (electron filling) and $h$. 
The other model parameters are taken at $t=-1$, $t_p=0.7$, $t_{z}=0.2$, $t_{sp}=0.3$, $\Gamma=0.8$, and $\theta=\pi/4$. 
As shown in Figs.~\ref{fig: Lattice_dat2}(a), \ref{fig: Lattice_dat2}(b), \ref{fig: Lattice_dat2}(e), and \ref{fig: Lattice_dat2}(f), $\langle G_z^{\rm (c)} \rangle$ and $\langle Q_0^{\rm (c)} \rangle$ shows a non-monotonic behavior against $\mu$. 
Meanwhile, $\langle M_0^{\rm (c)} \rangle$ and $\langle T_z^{\rm (c)} \rangle$ in Figs.~\ref{fig: Lattice_dat2}(c), \ref{fig: Lattice_dat2}(d), \ref{fig: Lattice_dat2}(g), and \ref{fig: Lattice_dat2}(h) tends to be larger close to the half-filling region, as often found in the itinerant electron model. 

\begin{acknowledgments}
This research was supported by JSPS KAKENHI Grants Numbers JP21H01037, JP22H04468, JP22H00101, JP22H01183, and by JST PRESTO (JPMJPR20L8). 
Parts of the numerical calculations were performed in the supercomputing systems in ISSP, the University of Tokyo.
\end{acknowledgments}

\bibliographystyle{apsrev}
\bibliography{ref}

\begin{thebibliography}{194}
\bibitem[{com()}]{comment_ET}
\bibinfo{note}{It is noted that the ferro-axial moment is present in the
  absence of $M_0$ and $T_z$ for the gray point groups $6/m 1'$, $6 1'$,
  $\bar{6} 1'$, $4/m 1'$, $4 1'$, $\bar{4} 1'$, $\bar{3} 1'$, $3 1'$, $2/m 1'$,
  $2 1'$, $m 1'$, $\bar{1} 1'$, and $1 1'$.}
\end{thebibliography}

\end{document}